\documentclass[journal]{IEEEtran}

\usepackage{amsmath,dsfont,yfonts,amssymb,epsfig,amsthm,bm,multirow, graphicx,color, mathrsfs, bbold}
\usepackage{stfloats}

\IEEEoverridecommandlockouts

\newtheorem{proposition}{Proposition}

\newtheorem{lemma}{Lemma}

\begin{document}

\title{Exploiting Channel Memory for Joint Estimation and Scheduling in Downlink Networks}

\author{{Wenzhuo Ouyang, Sugumar Murugesan, Atilla Eryilmaz, Ness B.
Shroff}
\vspace{-10pt}

\thanks{Wenzhuo Ouyang and Atilla Eryilmaz are with the Department of ECE, The Ohio State University (e-mails: ouyangw@ece.osu.edu, eryilmaz@ece.osu.edu). Sugumar Murugesan was with the Department of ECE, The Ohio State University and is currently with the School of ECEE, Arizona State University (e-mail: sugumar.murugesan@asu.edu). Ness B. Shroff holds a joint appointment in both the Department of ECE and the Department of CSE at
The Ohio State University (e-mail: shroff@ece.osu.edu). }
\thanks{This research was supported by NSF grants CAREER-CNS-0953515, CCF-0916664, CNS-0721236, CNS-0813000, DTRA Grant HDTRA 1-08-1-0016 and ARO MURI grant W911NF-08-1-0238.}}
\maketitle

\begin{abstract}
We address the problem of opportunistic multiuser scheduling in downlink networks with Markov-modeled outage channels. We consider the scenario in which the scheduler does not have full knowledge of the channel state information, but instead estimates the channel state information by exploiting the memory inherent in the Markov channels along with ARQ-styled feedback from the scheduled users. Opportunistic scheduling is optimized in two stages: (1) Channel estimation and rate adaptation to maximize the expected immediate rate of the scheduled user; (2) User scheduling, based on the optimized immediate rate, to maximize the overall long term sum-throughput of the downlink. The scheduling problem is a partially observable Markov decision process with the classic `exploitation vs exploration' trade-off that is difficult to quantify. We therefore study the problem in the framework of Restless Multi-armed Bandit Processes (RMBP) and perform a Whittle's indexability analysis. Whittle's indexability is traditionally known to be hard to establish and the index policy derived based on Whittle's indexability is known to have optimality properties in various settings. We show that the problem of downlink scheduling under imperfect channel state information is Whittle indexable and derive the Whittle's index policy in closed form. Via extensive numerical experiments, we show that the index policy has near-optimal performance.

Our work reveals that, under incomplete channel state information, exploiting channel memory for opportunistic scheduling can result in significant performance gains and that almost all of these gains can be realized using an easy-to-implement index policy.
\end{abstract}

\section{Introduction}
\label{sec:Intro}
The wireless channel is inherently time-varying and stochastic. It can be exploited for dynamically allocating resources to the network users, leading to the classic \textit{opportunistic scheduling} principle (e.g., \cite{Knopp}). Understandably, the success of opportunistic scheduling depends heavily on reliable knowledge of the instantaneous channel state information (CSI) at the scheduler. Many sophisticated scheduling strategies have been developed with provably optimal characteristics (e.g., \cite{Tse}-\cite{Atilla}) by assuming perfect CSI to be readily available, free of cost at the scheduler.

In realistic scenarios, however, perfect CSI is rarely, if ever, available and never cost-free, i.e., a non-trivial amount of network resource, that could otherwise be used for data transmission, must be spent in estimating the CSI \cite{Tse}. This calls for jointly designing channel estimation and opportunistic scheduling strategies -- an area that has recently received attention when the channel state is modeled by \textit{i.i.d.} processes across time (e.g., \cite{2stage}, \cite{Junshan2}). The \textit{i.i.d.} model has traditionally been a popular choice for researchers to abstract the fading channels, because of its simplicity and associated ease of analysis. On the other hand, this model fails to capture an important characteristics of the fading channels -- the time-correlation or the \textit{channel memory} \cite{Tse}.

In the presence of estimation cost, memory in the fading channels is an important resource that can be intelligently exploited for more  efficient, joint estimation and scheduling strategies. In this context, Markov channel models have been gaining popularity as realistic abstractions of fading channels with memory (e.g., \cite{Johnston}-\cite{ZhaoTWC}).


In this paper, we study joint channel estimation and scheduling using channel memory, in downlink networks. We model the downlink fading channels as two-state Markov Chains with \emph{non-zero achievable rate} in both states. The scheduling decision at any time instant is associated with two potentially contradicting objectives: (1) Immediate gains in throughput via data transmission to the scheduled user; (2) Exploration of the channel of a downlink user for more informed decisions and associated throughput gains in the future. This is the classic `exploitation vs exploration' trade-off often seen in sequential decision making problems  (e.g., \cite{clinical}, \cite{reinforce}). We model the joint estimation and scheduling problem as a Partially Observable Markov Decision Process (POMDP) and study the structure of the problem, by explicitly accounting for the estimation cost. Specifically, our contributions are as follows.
\vspace{3pt}

\noindent$\bullet$ We recast the POMDP scheduling problem as a Restless Multi-armed Bandit Process (RMBP) \cite{Whittle} and establish its \textit{Whittle's indexability} \cite{Whittle} in Section~\ref{sec:RMBP} and~\ref{sec:Indexability}. Even though Whittle's indexability is difficult to establish in general \cite{Glazebrook}, we have been able to show it in the context of our problem.

\noindent$\bullet$ Based on a Whittle's indexability condition, we explicitly characterize the Whittle's index policy for the scheduling problem in Section~\ref{sec:IndexPolicy}. Whittle's index policies are known to have optimality properties in various RMBP processes and have been shown to be easy to implement (e.g., \cite{Glazebrook}, \cite{Glazebrook2}).

\noindent$\bullet$ Using extensive numerical experiments, we demonstrate in Section~\ref{sec:numerical} that Whittle's index policy in our setting has near-optimal performance and that significant system-level gains can be realized by exploiting the channel memory for estimation and scheduling. Also, the Whittle's index policy we derive is of polynomial complexity in the number of downlink users (contrast this with the PSPACE-hard complexity of optimal POMDP solutions \cite{Tsitsiklis}).



Our setup significantly differs from related works (e.g., \cite{SM_IT} \cite{ZhaoTWC} \cite{Zhao_index}) in the following sense: In these works, the channels are modeled by ON-OFF Markov Chains with the OFF state corresponding to \textit{zero}-achievable rate of transmission. There, once a user is scheduled, there is no need to estimate the channel of that user, since it is optimal to transmit at the constant rate allowed by the ON state irrespective of the underlying state. In contrast, in our model, the achievable rate at the lower state is, in general, non-zero and any rate above this achievable rate leads to \emph{outage}. This extended model captures the realistic scenario when non-zero rates are possible with the use of sophisticated physical layer algorithms, even when the channel is bad. In this model, once a user is scheduled, the scheduler must estimate the channel of that user, with an associated cost, and adapt the transmission rate based on the estimate. The rate adaptation must balance between aggressive transmissions that lead to outage and conservative transmissions that lead to under-utilization of channels. The achievable rate expected from this process, in turn, influences the choice of the scheduled user. Thus the channel estimation and scheduling stages are tightly coupled, introducing several technical challenges to the problem, which we address in this paper.

\begin{figure}
\centering
\includegraphics[width=2.8in]{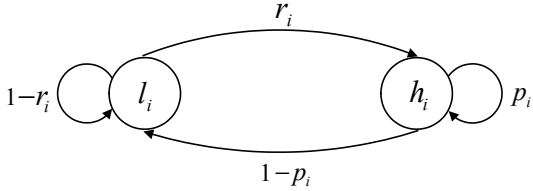}
\caption{A two state Markov Chain.}
\label{fig:chain}
\vspace{-5pt}
\end{figure}

\section{System Model and Problem Statement}
\label{sec:SysModel}

\subsection{Channel Model}
We consider a downlink system with one base station (BS) and $N$ users. Time is slotted with the time slots of all users synchronized. The channel between the BS and each user is modeled as a two-state Markov chain, i.e., the state of the channels remains static within each time slot and evolves across time slots according to Markov chain statistics. The Markov channels are assumed to be independent and, in general, non-identical across users. The state space of channel $C_i$ between the BS and user $i$ is given by $S_i=\{l_i,h_i\}$. Each state corresponds to a maximum allowable data rate. Specifically, if the channel is in state $l_i$, there exists a rate $\delta_i$, $0\le \delta_i<1$, such that data transmissions at rates below $\delta_i$ succeed and transmissions at rates above $\delta_i$ fail, i.e., outage occurs. Similarly, state $h_i$ corresponds to data rate $1$. Note that fixing the higher rate to be $1$ across \textit{all} users does not impose any loss of generality in our analysis. This will be evident as we proceed.

The Markovian channel model is illustrated in Fig.~\ref{fig:chain}. For user $i$, the two-state Markov channel is characterized by a $2\times 2$ probability transition matrix
\vspace{-2pt}
\begin{align}
P_i=\begin{bmatrix}
p_i&1-p_i\\
r_i&1-r_i\\
\end{bmatrix},\nonumber
\end{align}
where
\vspace{-9pt}
\begin{align}p_i&:= \textrm{prob$\big(C_i[t]{=}h_i \ \big | \ C_i[t{-}1]{=}h_i\big)$,}\nonumber \\
r_i&:= \textrm{prob$\big(C_i[t]{=}h_i \ \big | \ C_i[t{-}1]{=}l_i \big)$.}\nonumber
\end{align}
\begin{figure}
\centering
\includegraphics[width=2.7in]{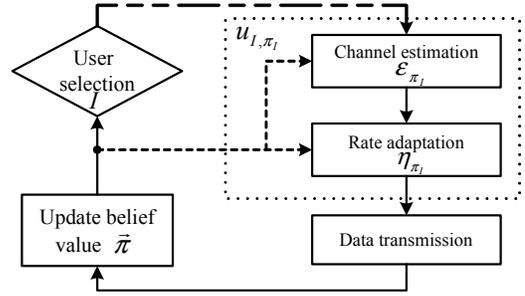}
\caption{Opportunistic scheduling with estimation and rate adaptation.}
\label{fig:process}
\end{figure}

\vspace{-15pt}
\subsection{Scheduling Model}
We adopt the one-hop interference model, i.e., in each time slot, only one user can be scheduled for data transmission. At the beginning of the slot, the scheduler does not have exact knowledge of the channel state of the downlink users. Instead, it maintains a belief value $\pi_i$ for channel $i$ which is the probability that $C_i$ is in state $h_i$ at the time. We will elaborate on the belief values soon. Using these belief values, the scheduler picks a user, estimates its current channel state and subsequently transmits data at a rate adapted to the channel state estimate -- all with an objective to maximize the overall sum-throughput of the downlink system. Specifically, in each slot, the scheduler jointly makes the following decisions: (1) Considering each user, the scheduler decides on the optimal channel estimator (that could involve the expenditure of network resources such as time, power, etc.) and rate adapter pair; (2) Based on the average rate of successful transmission promised for each user by the previous decision, the scheduler picks a user for channel estimation and subsequent data transmission. At the end of the slot, consistent with recent models (e.g., \cite{SM_IT} \cite{ZhaoTWC} \cite{Zhao_index}), the scheduled user sends back accurate information on the state of the Markov channel in that slot. This accurate feedback is, in turn, used by the scheduler to update its belief on the channels, based on the Markov channel statistics. Note that these belief values are sufficient statistics to the past scheduling decisions and feedback \cite{Sondik_thesis}. Using  $\varepsilon_{\pi}$ to denote  an arbitrary estimator and $\eta_{\pi}$ to denote an arbitrary rate adapter, as functions of the belief value $\pi$, the basic operation is summarized in Fig.~\ref{fig:process}. The scheduling problem can be formulated as a partially observable Markov decision process \cite{Sondik_thesis}, with the Markov channel states being the partially observable system states.

As noted in Section~\ref{sec:Intro}, the scheduling decision in each slot involves two objectives: data
transmission to the scheduled user and probing the channel of the scheduled user (through the accurate end-of-slot feedback). On one hand, the scheduler can transmit data to the user that promises the best achievable rate at the moment and hence realize immediate performance gains. On the other hand, the scheduler can schedule possibly another user and use the channel feedback from that user to gain a better understanding of the overall downlink system, which, in turn, could result in more informed future scheduling decisions with corresponding performance gains.


\subsection{Formal Problem Statement}
We now proceed to formally introduce the expected immediate reward. We let $\pi_i$ denote
the current belief value of the channel of user $i$, and let $u:=\{\varepsilon, \eta\}$ denote an arbitrary estimator and rate adapter pair. Recall from the discussion on the scheduling model that, at the end of the slot, the scheduled user sends back accurate feedback on its Markov channel state in that slot. With this setup, once a user is scheduled, the choice of the channel estimator and rate adapter pair does not affect the future paths of the scheduling process. Thus, within each slot, it is optimal to design this pair to maximize the expected rate (of successful transmission) of the user scheduled in that slot. Henceforth, in the language of POMDPs, we call this maximized rate the \textit{expected immediate reward}. We now proceed to formally introduce the expected immediate reward. We let $\pi_i$ denote the current belief value of the channel of user $i$. The optimal estimator and rate adapter pair, $u^*_{i, \pi_i}{=}\{\varepsilon^*_{i, \pi_i}, \eta^*_{i, \pi_i}\}$, for user $i$, when the belief is $\pi_i$, is given by
\begin{eqnarray}
\label{eq:OptimalPair}
u^*_{i,\pi_i}&=& \arg\max_{u} E_{C_i}[\gamma_i(C_i, u)],
\end{eqnarray}
where the quantity $\gamma_i(C_i, u)$ is the average rate of successful transmissions to user $i$ when the channel is in state $C_i$ and the estimator and rate adapter pair $u$ is deployed. The expectation in~(\ref{eq:OptimalPair}) is taken over the underlying channel state $C_i$, with distribution characterized by belief value $\pi_i$, i.e.,
\vspace{3pt}

\hspace{0.59in}$C_i=
\begin{cases}
h_i& \text{with probability $\pi_i$},\\
l_i& \text{with probability $1-\pi_i$}.
\end{cases}$
\vspace{4pt}

The expected immediate reward when user $i$ is scheduled is thus given by
\begin{align}
\label{eq:Rpi}
R_i(\pi_i)=E_{C_i}[\gamma_i(C_i,u^*_{i,\pi_i})].
\end{align}

\begin{figure}
\centering
\includegraphics[width=2.9in]{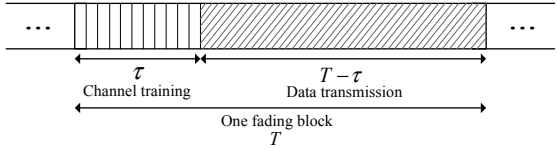}
\caption{A typical estimation - data transmission structure.}
\label{fig:block}
\end{figure}

Note that our model is very general in the sense that we do not restrict to any specific estimation, data transmission structure or to any specific class of estimators. A typical estimation, data transmission structure, corresponding to the estimator and rate adapter pair $u$ is illustrated in Fig.~\ref{fig:block}. Here a pilot-aided training\cite{Tse}-based estimation is performed for a fraction of the time slots followed by data transmission at an adapted rate in the rest of the time slots.

We now introduce the optimality equations for the scheduling problem. Let $\vec{\bm \pi}[t]=(\pi_1[t], \cdots, \pi_N[t])$ denote the vector of current belief values of the channels at the beginning of slot $t$. A stationary scheduling policy, $\Psi$, is a stationary mapping $\Psi: \vec{\bm \pi}\rightarrow I$ between the belief vector and the index of the user scheduled for data transmission in the current slot. Our performance metric is the infinite horizon, discounted sum-throughput of the downlink (henceforth simply the \textit{expected discounted reward} in the language of POMDPs), formally defined next.

For a stationary policy $\Psi$, the expected discounted reward under initial belief $\vec{\bm\pi}$ is given by
\begin{eqnarray}
\nonumber
V(\Psi,\vec{\bm\pi})&=&\sum_{t=0}^{\infty} \ \beta^{t} E_{\vec{\bm \pi}[t]}R_{I[t]=\Psi(\vec{\bm \pi}[t])}(\pi_{I[t]}[t])
\end{eqnarray}
where $\vec{\bm \pi}[t]$ is the belief vector in slot $t$, $\pi_{i}[t]$ denotes the belief value of user $i$ in slot $t$, $\vec{\bm\pi}[0]=\vec{\bm{\pi}}$, $I[t]$ denotes the index of the user scheduled in slot $t$. The discount factor $\beta\in[0,1)$ provides relative weighing between the immediately realizable rates and future rates. For any initial belief $\vec{\bm{\pi}}$, the optimal expected discounted reward, $V(\vec{\bm\pi})=\max_{\Psi}V(\Psi,\vec{\bm\pi})$, is given by the Bellman equation \cite{Bert}
\begin{eqnarray}
V(\vec{\bm \pi})&=&\max_{I} \{R_{I}(\pi_{I})+ \beta E_{\vec{\bm \pi}^+}[V(\vec{\bm \pi}^+)] \}. \nonumber
\end{eqnarray}

Here $\vec{\bm \pi}^+$ denotes the belief vector in the next slot when the current belief is $\vec{\bm \pi}$. The belief evolution $\vec{\bm \pi}\rightarrow\vec{\bm \pi}^+$ proceeds as follows:

\begin{align}
\pi_i^+=
\begin{cases}
p_i& \text{if $I=i$ and $C_i=h_i$}\\
r_i& \text{if $I=i$ and $C_i= l_i$} ,\\
Q_i(\pi_i)& \text{if $I\neq i$}
\end{cases}
\end{align}

\noindent where $Q_i(x)=x p_i+(1-x)r_i$ is the belief evolution operator for user $i$ when it is not scheduled in the current slot. A stationary scheduling policy $\Psi^*$ is optimal if and only if $V(\Psi^*,\vec{\bm\pi})=V(\vec{\bm\pi})$ for all $\vec{\bm\pi}$ \cite{Bert}.

In the introduction, we briefly contrasted our setup with those in \cite{SM_IT}\cite{ZhaoTWC}\cite{Zhao_index}. We provide a rigorous comparison here. The works \cite{SM_IT}\cite{ZhaoTWC}\cite{Zhao_index} studied opportunistic scheduling with the channels modeled by ON-OFF Markov chains. In these works, the lower state is an `OFF' state, i.e., it does not allow transmission at any non-zero data rate. Contrast this with our model where, at the lower state $l_i$, a possibly non-zero rate $\delta_i$ is achievable and outage occurs at any rate above $\delta_i$. We now further explain how these two models are fundamentally different.

\noindent $\bullet$ In the ON-OFF channel model, the scheduler does not need a channel estimator and rate adapter pair. The scheduler can aggressively transmit at rate $1$, since it has nothing to gain by transmitting at a lower rate -- a direct consequence of the `OFF' nature of the lower state. On the other hand, transmitting at a rate lesser than $1$ can lead to losses due to under-utilization of the channel.
\vspace{2pt}

\noindent $\bullet$ In contrast, in our model, when $\delta>0$, the scheduler must strike a balance between aggressive and conservative rates of transmission. An aggressive strategy (transmit at rate $1$) can lead to losses due to outages, while a conservative strategy can lead to losses due to under-utilization of the channel. This underscores the importance of the knowledge of the underlying channel state and, therefore, the need for intelligent estimation and rate adaptation mechanisms.
\vspace{2pt}

\noindent $\bullet$ As a direct consequence of the preceding arguments, the expected immediate reward in our model is not a trivial $\delta$-shift of the expected immediate reward when the rates supported by the channel states are $0$ and $1-\delta$. Formally,
\begin{align}
R^{\{\delta,1\}}(\pi) \neq R^{\{0,1-\delta\}}(\pi)+\delta=(1-\delta)\pi+\delta, \nonumber
\end{align}
where $R^{\{x,y\}}(\pi)$ is the immediate reward when the channel state space is $\{x,y\}$ and belief value of the scheduled user is $\pi$. In fact, it can be shown that (in Lemma~\ref{lemma:convexRpi})
\begin{align}
R^{\{\delta,1\}}(\pi) \le R^{\{0,1-\delta\}}(\pi)+\delta=(1-\delta)\pi+\delta. \nonumber
\end{align}
%

We believe that, our channel model, in contrast to the ON-OFF model, better captures realistic communication channels where, using appropriate physical layer algorithms, it is possible to transmit at a non-zero rate even at the lowest state of the channel model and the same physical layer algorithms may impose outage behavior when this allowed rate is exceeded.

\section{Optimal Expected Transmission Rate -- Structural Properties}

In this section, we study the structural properties of the expected immediate reward, $R_i(\pi_i)$, defined in Equation~(\ref{eq:Rpi}). These properties will be crucial for our analysis in subsequent sections. For notational convenience, we will drop the suffix $i$ in the rest of this section.


\begin{lemma}
\label{lemma:convexRpi}
The expected immediate reward $R(\pi)$ has the following properties:\\
(a) $R(\pi)$ is convex and increasing in $\pi$ for $\pi \in [0,1]$\\
(b) $R(\pi)$ is bounded as follows:
\begin{align}
\label{eq:Rbounds}
\max \{\delta, \pi \} \leq R(\pi)\leq (1-\delta)\pi+\delta.
\end{align}
\end{lemma}


\noindent \textbf{Proof:} Let $U^*$ be the set of optimal estimator and rate adapter pairs for all $\pi\in[0,1]$, i.e., $U^*=\{u_{\pi}^*, \pi \in [0,1] \}$. The expected immediate reward, provided in~(\ref{eq:Rpi}), can now be rewritten as
\begin{align}
R(\pi)&=\max_{u \in U^*} E_C[\gamma(C,u)] \nonumber\\
&=\max_{u \in U^*}[\pi \gamma(h,u) + (1-\pi) \gamma(l,u)],\nonumber
\end{align}
where $\gamma(s,u)$ denotes the average rate of successful transmission when the channel state is $s\in \{l,h\}$. Note that, for fixed $u$, the average rate $\pi \gamma(h,u) + (1-\pi) \gamma(l,u)$ is linear in $\pi$. Thus, $R(\pi)$ is given as a point-wise maximum over a family of linear functions, which is convex \cite{Boyd}. $R(\pi)$ is therefore convex in $\pi$, establishing the convexity statement in (a).

We next proceed to derive the bounds to $R(\pi)$. From Equation~(\ref{eq:Rpi}),
\begin{eqnarray}
R(\pi)=\max_{u} E_{C} [\gamma(C,u)] \geq \max_{\{u: u=\{\eta\}\}} E_C [\gamma(C,u)] \nonumber
\end{eqnarray}

\noindent where $\{ u:u=\{\eta\} \}$ indicates that we are considering rate adaptation without channel estimation. This explains the last inequality. Note that without the estimator, the rate adaptation is solely a function of the belief value $\pi$. Thus, the average rate achieved under the rate adapter, conditioned on the underlying channel state, can be expressed simply by indicator functions, as seen below:
\begin{eqnarray}
\lefteqn{\max_{\{u: u=\{\eta\} \}} E_C [\gamma(C,u)]}\nonumber\\
&=&\max_{\eta} \  [P(C=l) \eta \cdot \bm  1(\eta\leq \delta)+P(C=h) \eta \cdot \bm  1(\eta\leq 1)] \nonumber\\
&=&\max_{\eta}  \eta [P(C=l)  \cdot \bm  1(\eta\leq \delta)+P(C=h) \cdot \bm  1(\eta\leq 1)] \nonumber\\
&=&\max \ \{\delta, \pi\}. \nonumber
\end{eqnarray}

This establishes the lower bound in (b).

The upper bound in (b) corresponds to the expected immediate reward when \textit{full} channel state information is available at the scheduler.

It is clear from the upper and lower bounds that $\delta {\leq} R(\pi)\leq 1$. Note that when $\pi{=}0$ or $\pi{=}1$, there is no uncertainty in the channel, hence $R(0){=}\delta$ and $R(1){=}1$. Using these properties, along with the convexity property of $R(\pi)$, we see that $R(\pi)$ is monotonically increasing in $\pi$, establishing the monotonicity of (a). The lemma thus follows.  $\hfill \blacksquare$

\vspace{6pt}


\noindent \textbf{Remark:}
Here we present some insights into the effect of the non-zero rate $\delta$ on the channel estimation and rate adaptation mechanisms by studying the upper and lower bounds to $R(\pi)$ provided in Lemma~\ref{lemma:convexRpi}. The upper bound essentially corresponds to the case when perfect channel state information is available at the scheduler at the beginning of each slot. Here, no channel estimation and rate adaptation is necessary. The lower bound, on the other hand, corresponds to the case when the channel estimation stage is eliminated and rate adaptation is performed solely based on the belief value $\pi$ of the scheduled user.

\begin{figure}
\centering
\includegraphics[width=3.2in]{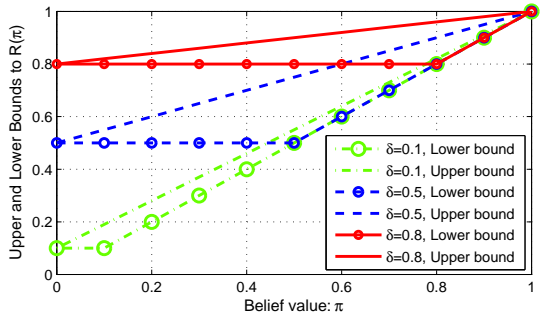}
\caption{Upper and lower bounds to the average rate of successful transmission.}
\vspace{-12pt}
\label{fig:bounds}
\vspace{-5pt}
\end{figure}

Fig.~\ref{fig:bounds} plots the lower and upper bounds to $R(\pi)$ for different values of $\delta$. Note that the lower bound approaches the upper bound in both directions, i.e., when $\delta \rightarrow 0$ or when $\delta\rightarrow 1$. This behavior can be explained as follows: (1) $\delta\rightarrow 1$ essentially means that the states of the Markov channel move closer to each other. This progressively reduces the channel uncertainty and hence the need for channel estimation (and, consequently, rate adaptation), essentially bringing the bounds closer. (2) As $\delta\rightarrow 0$, the channel uncertainty increases. At the same time, the impact of the channel estimator and rate adapter pair decreases. This is because, as $\delta\rightarrow 0$, the loss in immediate reward due to outage (transmitting at $1$ when channel is in state $\delta$) is less severe than the loss due to under-utilization of the channel (transmitting at rate $\delta$ when the channel is in state $1$), essentially making it optimal for the rate adaptation scheme to be progressively more aggressive (transmit at rate $1$). Thus channel estimation loses its significance as $\delta\rightarrow 0$. This brings the bounds closer as $\delta\rightarrow 0$.

It can be verified that the separation between the lower and upper bounds is at its peak when $\delta=0.5$.
This, along with the preceding discussion, indicates the potential for rate improvement when intelligent channel estimation and rate adaptation is performed under moderate values of $\delta$.

\section{Restless Multi-Armed Bandit Processes, Whittle's Indexability and Index Policies}
\label{sec:RMBP}

A direct analysis of the downlink scheduling problem appears difficult due to the complex nature of the
`exploitation vs exploration' tradeoff. We therefore establish a connection between the scheduling problem and the Restless Multiarmed Bandit Processes (RMBP) \cite{Whittle} and make use of the established theory behind RMBP in our analysis. We briefly overview RMBPs and the associated theory of Whittle's indexability in this section.

RMBPs are defined as a family of sequential dynamic resource allocation problems in the presence of several competing, independently evolving projects. In RMBPs, a subset of the competing projects are served in each slot. The states of all the projects in the system stochastically evolve in time based on the current state of the projects and the action taken. Once a project is served, a reward dependent on the states of the served projects and the action taken is accrued by the controller. Hence, the RMBPs are characterized by a fundamental tradeoff between decisions guaranteeing high immediate rewards versus those that sacrifice immediate rewards for better future rewards. Solutions to RMBPs are, in general, known to be PSPACE-hard \cite{Tsitsiklis}.

Under an \textit{average} constraint on the number of projects scheduled per slot, a low complexity index policy developed by Whittle \cite{Whittle}, commonly known as Whittle's index policy, is optimal.
Under stringent constraint on the number of users scheduled per slot, Whittle's index policy may not exist and if it does exist, its optimality properties are, in general, lost. However, Whittle's index policies, upon existence, are known to have near optimal performance in various RMBPs (e.g., \cite{Glazebrook} \cite{Glazebrook2}). For an RMBP, Whittle's index policy exists if and only if the RMBP satisfies a condition known as \emph{Whittle's indexability} \cite{Whittle}, defined next.

Consider the following setup: for each project $P$ in the system, consider a virtual system where, in each slot, the controller must make one of two decisions: (1) Serve project $P$ and accrue an immediate reward that is a function of the state of the project. This reward structure reflects the one in the original RMBP for project $P$. (2) Do not serve project $P$, i.e., stay passive and accrue an immediate reward for passivity $\omega$. The state of the project $P$ evolves in the same fashion as it would in the original RMBP, as a function of its current state and current action (whether $P$ is served or not in the current state). Let $D(\omega)$ be the set of states of project $P$ in which it is optimal to stay passive, where optimality is defined based on the infinite horizon net reward.

\textit{Project $P$ is Whittle indexable if and only if \textit{as $\omega$ increases from $-\infty$ to $\infty$, the set $D(\omega)$ monotonically expands from $\emptyset$ to $S$, the state space of project $P$}. The RMBP is Whittle indexable if and only if \textit{all} the projects in the RMBP are Whittle indexable.}

For each state, $s$, of a project, Whittle's index, $W(s)$, is given by the value of $\omega$ in which the net reward after both the active and passive decisions are the same in the $\omega$-subsidized virtual system. The notion of indexability gives a consistent ordering of states with respect to the indices. For instance, if $W(s_1){>}W(s_2)$ and if it is optimal to serve the project at state $s_1$, then it is optimal to serve the project at $s_2$. This natural ordering of states based on indices renders the near-optimality properties to Whittle's index policy (e.g., \cite{Glazebrook}, \cite{Glazebrook2}).

The downlink scheduling problem we have considered is in fact an RMBP process. Here, each downlink user, along with the belief value of its channel, corresponds to a project in the RMBP, and the project is served when the corresponding user is scheduled for data transmission. Now, referring to our earlier discussion on the RMBPs, we see that Whittle's index policy is very attractive from an optimality point of view. The attractiveness of the index policy can be attributed to the natural ordering of states (and hence projects) based on indices, as guaranteed by Whittle's indexability. In the rest of the paper, we establish that this advantage carries over to the downlink scheduling problem at hand. As a first step in this direction, in the next section, we study the scheduling problem in Whittle's indexability framework and show that the downlink scheduling problem is, in fact, Whittle indexable.

\section{Whittle's Indexability Analysis of the Downlink Scheduling Problem}
\label{sec:Indexability}



In this section, we study the Whittle's indexability of our joint scheduling and estimation problem. To that end, we first describe the downlink scheduling setup:

At the beginning of each slot, based on the current belief value $\pi$ (we drop the user index $i$ in this section since only one user is considered throughout), the scheduler takes one of two possible actions: schedules data transmission to the user (action $a=1$) or stays idle ($a=0$). Upon an idle decision, a \emph{subsidy} of $\omega$ is obtained. Otherwise, optimal channel estimation and rate adaptation is carried out, with a reward equal to $R(\pi)$ (consistent with the immediate reward seen in previous sections). The belief value is updated based on the action taken and feedback from the user (upon transmit decision). This belief update is consistent with that in the Section~\ref{sec:SysModel}. The optimal scheduling policy (henceforth, the $\omega$-subsidy policy) maximizes the infinite horizon discounted reward, parameterized by $\omega$. The optimal infinite horizon discounted reward is given by the Bellman equation \cite{Bert}
\begin{align}
\label{eq:Bellman}
V_{\omega}(\pi)=\max  \{&\big[R(\pi)+\beta \big(\pi
V_{\omega}(p)+(1-\pi) V_{\omega}(r) \big)\big], \nonumber\\
&\hspace{3pt}\big[\omega +\beta V_{\omega} \big(Q(\pi) \big)\big]\},
\end{align}
where, recall from Section~\ref{sec:SysModel}, $Q(\pi)$ is the evolution of the belief value when the user is not scheduled. The first quantity inside the $\max$ operator corresponds to the infinite horizon reward when a \textit{transmit} decision is made in the current slot and optimal decisions are made in the future slot. The second element corresponds to \textit{idle} decision in the current slot and optimal decisions in all future slots.

We note that the indexability analysis in the rest of this section bears similarities to that in \cite{Zhao_index}, where the authors studied indexability of a sequential resource allocation problem in a cognitive radio setting. This problem is mathematically equivalent to our downlink scheduling problem when $\delta=0$. We have already discussed in detail (in Section~\ref{sec:SysModel}) that the structure of the immediate reward $R(\pi)$ when $\delta>0$ is very different than when $\delta=0$, due to the need for channel estimation and rate adaptation in the former case. Consequently, in the Whittle's indexability setup, the infinite horizon discounted reward $V_{\omega}(\pi)$ in our problem is different (and more general) than that in \cite{Zhao_index}, underscoring the significance of our results.

As a crucial preparatory result, we now proceed to show that the $\omega$-subsidy policy is \textit{thresholdable}.

\subsection{Thresholdability of the $\omega$-subsidy policy}

We first record our result on the convexity property of the infinite horizon discounted reward, $V_{\omega}(\pi)$, of~(\ref{eq:Bellman}) in the following proposition.


\begin{proposition}
\label{prop:Vpiconvex}
The infinite horizon discounted reward, $V_{\omega}(\pi)$ is convex in $\pi \in [0,1]$.
\end{proposition}

\noindent \textbf{Proof:} We first consider the discounted reward
for finite horizon $\omega$-subsidy problem. We let $\nu^{1}(\pi)=R(\pi)$ and $\nu^{0}(\pi)=\omega$ represent the immediate reward corresponding to active and idle decisions, respectively. The reward function associated with $M$-stage finite horizon process is expressed as
\begin{align}
\widetilde{V}_{M}(\pi[0])=\max_{ \begin{subarray}{c} a[t], \\ t=0,{\ldots},M-1 \end{subarray}} E\Big[\sum_{t=0}^{M} \beta^{t}
\nu^{a[t]}(\pi[t]) \Big| \pi[0]\Big] \nonumber
\end{align}

Let $\widehat{V}_{\omega,t}(\pi)$ be the reward at time $t$ with belief value $\pi[t]=\pi$. Hence $\widetilde{V}_{M}(\pi[0])=\widehat{V}_{\omega,0}(\pi[0])$ and the last stage value function $\widehat{V}_{\omega,M-1}(\pi[M-1])$ is given by
\begin{align}
\widehat{V}_{\omega,M-1}(\pi[M-1])&=\max_{a[M-1]} \ \{\nu^{a[M-1]}(\pi[M-1]) \} \nonumber \\
&=\max \{ \omega, R(\pi[M-1]) \}. \nonumber
\end{align}

Therefore, $\widehat{V}_{\omega,M-1}(\pi)$ is convex with $\pi$ since it is the maximum of a constant and a convex function. For any time $0\leq t< M-1$, the Bellman (\cite{Bert}) equation can be written as
\begin{align}
\widehat{V}_{\omega,t}(\pi[t])= \max \{\widehat{V}^0_{\omega,t}(\pi[t]), \widehat{V}^1_{\omega,t}(\pi[t])\}. \nonumber
\end{align}
where
\begin{align}
\widehat{V}^{0}_{\omega,t}(\pi)&{=}\omega {+}\beta \widehat{V}_{\omega,t+1} \big(Q(\pi) \big), \label{eq:V0}\\
\widehat{V}^{1}_{\omega,t}(\pi)&{=} R(\pi){+}\beta \big(\pi \widehat{V}_{\omega,t+1}(p)+(1-\pi) \widehat{V}_{\omega,t+1}(r) \big).\label{eq:V1}
\end{align}

Suppose now $\widehat{V}_{\omega, t+1}(\pi)$ is convex with $\pi$. If $a[t]=1$, it is clear from~(\ref{eq:V1}) that $\widehat{V}^{1}_{\omega, t}(\pi)$ is convex function of $\pi$ since it is a summation of a convex function and a linear function of $\pi$. If $a[t]=0$, $\widehat{V}^{0}_{\omega, t}(\pi)$, expressed in~(\ref{eq:V0}), is also a convex function, because composition of convex function $\widehat{V}_{\omega,t+1}(\cdot)$ and linear function $Q(\pi)$ is convex \cite{Boyd}. Therefore $\widehat{V}_{\omega, t}(\pi)$ is convex with $\pi$ as maximum of two convex functions. By induction, the convexity of $V_{\omega, 0}(\pi)$ is thus established.
\vspace{4pt}

Since $\widetilde{V}_{M}(\pi)=V_{\omega, 0}(\pi)$, $\widetilde{V}_{M}(\pi)$ is convex with $\pi$. For discounted problem with bounded reward per slot, the infinite horizon reward is the limit of of finite horizon reward (\cite{Bert}). Therefore $V_{\omega}(\pi)= \lim_{M\rightarrow \infty} V_{\omega, M}(\pi)$. Upon point-wise convergence, point-wise limit of convex functions is convex \cite{Boyd}. Hence $V_{\omega}(\pi)$ is a convex function of $\pi$.$\hfill \blacksquare$

\newcounter{TempEqCnt}
\setcounter{TempEqCnt}{\value{equation}}
\setcounter{equation}{10}

\begin{figure*}[hb]
\hrulefill
\begin{align}
\label{eq: Theta}
\Theta=\frac{(1-\beta^{L(r,\pi^*(\omega))})\omega+(1-\beta)
\beta^{L(r, \pi^*(\omega))}[R(Q^{L(r,\pi^*(\omega))}(r))+\beta
Q^{L(r,\pi^*(\omega))}(r) V_{\omega}(p)]}{(1-\beta)[1-\beta^{L(r,\pi^*(\omega))+1}
(1-Q^{L(r,\pi^*(\omega))}(r))]}
\end{align}
\vspace{-0.2in}
\end{figure*}

\setcounter{equation}{\value{TempEqCnt}}

\vspace{3pt}

In the next proposition, we show that the optimal $\omega$-subsidy policy is a threshold policy.

\begin{proposition}
\label{prop:thresd}
The optimal $\omega$-subsidy policy is thresholdable in the belief space $\pi$. Specifically, there exists a threshold $\pi^*(\omega)$ such that the optimal action $a$ is $1$ if the current belief $\pi> \pi^*(\omega)$ and the optimal action $a$ is $0$, otherwise. The value of the threshold $\pi^*(\omega)$ depends on the subsidy $\omega$, partially characterized below.


\vspace{8pt}
\noindent(i) If $\omega\geq 1$, $\pi^*(\omega)=1$;\\
(ii) If $\omega\leq \delta$, $\pi^*(\omega)=\kappa$ for some arbitrary $\kappa<0$;\\
(iii) If $\delta < \omega < 1$, $\pi^*(\omega)$ takes value within interval $(0,1)$.
\end{proposition}

\noindent \textbf{Proof:} Consider the Bellman equation~(\ref{eq:Bellman}), let $V_{\omega}^1(\pi)$ be the reward corresponding to transmit decision and $V_{\omega}^0(\pi)$ be the reward corresponding to idle decision, i.e.,
\begin{align}
V_{\omega}^1(\pi)&=R(\pi)+\beta \big(\pi
V_{\omega}(p)+(1-\pi) V_{\omega}(r) \big), \label{eq:V1_Bellman}\\
V_{\omega}^0(\pi)&=\omega+ \beta V_{\omega} \big(Q(\pi) \big)=\omega+ \beta V_{\omega} \big(\pi p+(1-\pi)r \big). \label{eq:V0_Bellman}
\end{align}

It is clear from the Bellman equation~(\ref{eq:Bellman}) that the optimal action depends on the relationship between $V_{\omega}^1(\pi)$ and $V_{\omega}^0(\pi)$, presented as follows.
\vspace{5pt}

\noindent Case (i). If $\omega\geq1$, since $R(\pi)\leq1$, in each slot, the immediate reward for being idle always dominates the reward for being active. Hence it will be optimal to always stay idle. We can thus set the threshold to 1.

\noindent Case (ii). If $\omega \leq \delta$, then for any $\pi \in
[0,1]$, we have
\begin{align}
V_{\omega}^0(\pi)=&\omega+\beta V_{\omega} (\pi p+(1-\pi) r)\nonumber \\
\leq& R(\pi) +\beta (\pi V_{\omega}(p)+(1-\pi)
V_{\omega}(r)), \nonumber\\
=& V_{\omega}^1(\pi), \nonumber
\end{align}
where the inequality is due to $\delta {\leq} R(\pi)$ along with Jensen's inequality \cite{Boyd} due to the convexity of $V_{\omega}(\pi)$ from Proposition 2. Hence, it is optimal to stay active. Consistent with the threshold definition, we can set $\pi^*(\omega)=\kappa$ for any $\kappa<0$.

\noindent Case (iii). If $\delta < w < 1$, then at the extreme values of belief,
\begin{align}
V_{\omega}^0(0)&=\omega+\beta V_{\omega} (r) > \delta + \beta V_{\omega} (r)= V_{\omega}^1 (0) \nonumber \\
V_{\omega}^0(1)&=\omega+\beta V_{\omega} (p) < 1 + \beta V_{\omega} (p)= V_{\omega}^1(1) \nonumber
\end{align}

Note that the relationship of $V_{\omega}^0(\pi)$ and $V_{\omega}^1(\pi)$ is reversed at the end points $0$ and $1$, and they are both convex functions of $\pi$. Thus, there must exist a threshold $\pi^*(\omega)$ within $(0,1)$ such that $a$ equals 1 whenever $\pi>\pi^*(\omega)$. $\hfill \blacksquare$
\vspace{4pt}

\subsection{Whittle's Indexability of Downlink Scheduling}

Having established that the $\omega$-subsidy policy is thresholdable in Proposition~\ref{prop:thresd}, Whittle's indexability, defined in Section~\ref{sec:RMBP}, is re-interpreted for the downlink scheduling problem as follows: the downlink scheduling problem is Whittle indexable if the threshold boundary $\pi^*(\omega)$ is monotonically increasing with subsidy $\omega$.

Using our discussion in Section~\ref{sec:RMBP}, the index of the belief value $\pi$, i.e., $W(\pi)$ is the infimum value of the subsidy $\omega$ such that it is optimal to stay idle, i.e.,
\begin{align}
W(\pi)&= \inf \{ \omega: V^0_{\omega}(\pi) \geq V^1_{\omega}(\pi) \}\nonumber \\
&=\inf \{ \omega: \pi^*(\omega)=\pi \}. \label{eq:Index_val}
\end{align}
%

\setcounter{equation}{11}

To establish indexability, we need to investigate the infinite horizon discounted reward $V_{\omega}(\pi)$, given by~(\ref{eq:Bellman}). We can observe from~(\ref{eq:Bellman}) that given the value of $V_{\omega}(p)$ and $V_{\omega}(r)$, $V_{\omega}(\pi)$ can be calculated for all $\pi \in [0,1]$. Let $\pi^0$ denote the steady state probability of being in state $h$. The next lemma provides a closed form expression for $V_{\omega}(p)$ and $V_{\omega}(r)$ and is critical to the proof of indexability.


\begin{lemma} The discounted rewards $ V_{\omega}(p)$ and $V_{\omega}(r)$ can be expressed as:
\label{lemma:VpVr}
\vspace{2pt}

\noindent Case 1: $p > r$ (positive correlation)

\vspace{4pt}
\noindent $V_{\omega}(p){=}
\begin{cases}
\frac{R(p)+\beta(1-p)V_{\omega}(r)}{1-\beta p} &\text{\hspace{0.646in}if $\pi^*(\omega) < p$}\\
\frac{\omega}{1-\beta} &\text{\hspace{0.646in}if $\pi^*(\omega) \geq p$}\\
\end{cases}$\\
\vspace{3pt}

\noindent $V_{\omega}(r){=}
\begin{cases}
\sum_{k=0}^{\infty} \beta^k R(\frac{r{-}(p{-}r)^{k{+}1}r}{1{+}r{-}p}) &\text{\hspace{0.2in}if $\pi^*(\omega) {<}  r$}\\
\Theta &\text{\hspace{0.2in}if $r {\leq} \pi^*(\omega) {<} \pi^0$}\\
\frac{\omega}{1-\beta} &\text{\hspace{0.2in}if $\pi^*(\omega) \geq \pi^0$}\\
\end{cases}$\\

\vspace{6pt}
\noindent Case 2: $p \leq r$ (negative correlation)

\vspace{3pt}
\hspace{-10pt} $V_{\omega}(p){=}\hspace{-3pt}
\begin{cases}
\sum_{k{=}0}^{\infty} \beta^k
R(\frac{r{+}(p{-}r)^{k{+}1}(1{-}p)}{1{+}r{-}p}) \hspace{-7pt} & \text{\hspace{0.09in}if $\pi^*(\omega) {<}  p$}\\
\frac{\omega{+}\beta R(Q(p)){+}\beta^2 (1{-}Q(p))V_{\omega} (r)}{1{-}\beta^2 Q(p)}
\hspace{-7pt} & \text{\hspace{0.09in}if $p {\leq} \pi^*(\omega) {<} Q(p)$}\\
\frac{\omega}{1-\beta} \hspace{-7pt} & \text{\hspace{0.09in}if $\pi^*(\omega) {\geq} Q(p)$}\\
\end{cases}$
\vspace{4pt}

\hspace{-10pt} $V_{\omega}(r){=}\hspace{-3pt}
\begin{cases}
\frac{R(r)+\beta r V_{\omega}(p)}{1-\beta(1-r)} & \hspace{0.91in}\text{if $\pi^*(\omega) {<} r$}\\
\frac{\omega}{1-\beta} & \hspace{0.91in}\text{if $\pi^*(\omega) {\geq} r$}\\
\end{cases}$
\vspace{3pt}

The expression of $\Theta$ is given by Equation~(\ref{eq: Theta}), where $Q^n$ denotes $n^{th}$ iteration of $Q$ and $L(\pi, \pi^*(\omega))$ is a function of $\pi$ and $ \pi^*(\omega)$. Their expressions are given in Appendix~\ref{appen:VpVr}. From the above expressions, the closed form $V_{\omega}(p)$ and $V_{\omega}(r)$ can be readily obtained. The explicit expression is space-consuming and therefore is moved to Appendix~\ref{appen:VpVr}.
\end{lemma}
\vspace{3pt}

\noindent\textbf{Proof:} The derivation of $V_{\omega}(p)$ and $V_{\omega}(r)$ follows from substituting $p$ and $r$ in Equation~(\ref{eq:Bellman}). Together with the expression of $Q(\pi)$ given by in Section~\ref{sec:SysModel}, the expression of $V_{\omega}(p)$ and $V_{\omega}(r)$ can be obtained. For details, please refer to Appendix~\ref{appen:VpVr}. $\hfill \blacksquare$
\vspace{4pt}

We note that the value function expression depends on the correlation type of the Markov chain, because the transition function $Q(\pi)$ given in Section~\ref{sec:SysModel} will behave differently with the correlation type of the chain.

The closed form expression of the value function given by the previous lemma serves as a useful tool for us to establish indexability, which is given by the next proposition.


\begin{proposition}
\label{prop:Indexability}
The threshold value is strictly increasing with $\omega$. Therefore, the problem is Whittle indexable.
\end{proposition}

\noindent \textbf{Proof:} The proof of indexability follows the lines of \cite{Zhao_index}. Details are provided in Appendix~\ref{appen:Indexability}. $\hfill \blacksquare$

\section{Whittle's Index Policy}
\label{sec:IndexPolicy}
\vspace{4pt}

\section{Whittle's Index Policy and Numerical Performance Analysis}


In this section, we explicitly characterize Whittle's index policy for the downlink scheduling problem. For user $i$, let $\pi^0_i$ denote the steady state probability of being in state $h_i$, and let $V_{i,\omega}(\pi_i)$ denote the reward function for its $\omega$-subsidy problem in~(\ref{eq:Bellman}). We first characterize the Whittle's index as follows.


\begin{proposition}
\label{prop:Index_val}
For user $i$, the index value at state $\pi_i$, i.e., $W_i(\pi_i)$ is characterized as follows,
\vspace{5pt}

\noindent Case 1. Positively correlated channel ($p_i > r_i$)\\

{\small
\hspace{-15pt}$W_i(\pi_i){=}\hspace{-3pt}
\begin{cases}
R_i(\pi_i) \hspace{1.79in} \text{if \hspace{19pt} $\pi_i \geq p_i$}\\
\frac{\beta \pi_i R_i(p_i) + (1-\beta p_i) R_i(\pi_i)
}{1 + \beta \pi_i -\beta p_i} \hspace{0.89in} \text{if\hspace{1pt} $\pi_i^0 \leq \pi_i < p_i$}\\
[R_i(\pi_i){-}\beta R_i(Q_i(\pi_i))]{+}\beta[\pi_i {-}\beta Q_i(\pi_i)]V_{i,W_i(\pi_i)}(p_i)\\
{+}\beta[(1{-}\pi_i){-}\beta (1{-}Q_i(\pi_i))]V_{i,W_i(\pi_i)}(r_i) \hspace{0.1in} \text{if \hspace{0.22in} $\pi_i < \pi_i^0$}\\
\end{cases}$
}
\vspace{4pt}

\noindent Case 2. Negatively correlated channel ($p_i\leq r_i$)\\

{\small
\hspace{-15pt}$W_i(\pi_i){=}\hspace{-3pt}
\begin{cases}
R_i(\pi_i) \hspace{1.713in} \text{if \hspace{0.38in}$\pi_i \geq r_i$}\\
\frac{(1-\beta)[R_i(\pi_i)+\beta(1-\pi_i)V_{i,W_i(\pi_i)}(r_i)]}{1-\beta \pi_i} \hspace{0.32in} \text{if \ $Q_i(p_i) {\leq} \pi_i {<} r_i$}\\
(1{-}\beta)\big[R_i(\pi_i){+}\beta[\pi_i V_{i,W_i(\pi_i)}(p_i){+}(1{-}\pi_i)V_{i,W_i(\pi_i)}(r_i)]\big]\\
\hspace{2.07in}\text{if $\pi_i^0 {\leq} \pi_i {<} Q_i(p_i)$}\\
[R_i(\pi_i){-}\beta R_i(Q_i(\pi_i))]{+}\beta[\pi_i {-}\beta Q_i(\pi_i)]V_{i,W_i(\pi_i)}(p_i)\\
{+}\beta [(1{-}\pi_i){-}\beta (1{-}Q_i(\pi_i))]V_{i,W_i(\pi_i)}(r_i) \hspace{0.33in} \text{if
$\pi_i < \pi_i^0$}\\
\end{cases}$
}
\end{proposition}
\vspace{5pt}

\noindent \textbf{Proof:} The derivation of the index value follows from substituting the expression of $V_{i,\omega_i}(p_i)$ and $V_{i,\omega}(r_i)$ (given in Lemma~\ref{lemma:VpVr}) into Equation~(\ref{eq:Bellman}). Details of the proof are provided in Appendix~\ref{appen:Index_val}. $\hfill \blacksquare$
\vspace{3pt}

\noindent \textbf{Remark:} Notice that Proposition~\ref{prop:Index_val} does not give the closed form expression for $W_i(\pi_i)$. However, since the closed form expression of the value function $V_{i,W_i(\pi_i)}(p_i)$ and $V_{i,W_i(\pi_i)}(r_i)$ are derived in Lemma~\ref{lemma:VpVr}, closed form expressions of $W_i(\pi_i)$ can be easily calculated and is given in Appendix~\ref{appen:Index_val}. We now introduce Whittle's index policy.
\vspace{5pt}

\noindent \textbf{Whittle's Index Policy:} \emph{In each slot, with belief values $\pi_1,\ldots, \pi_N$, the user $I$ with the highest index value $W_i(\pi_i)$ is scheduled for transmission, i.e., $I=\arg\max_i W_i(\pi_i)$.}
\vspace{8pt}

\begin{figure}
\centering
\includegraphics[width=2.6in]{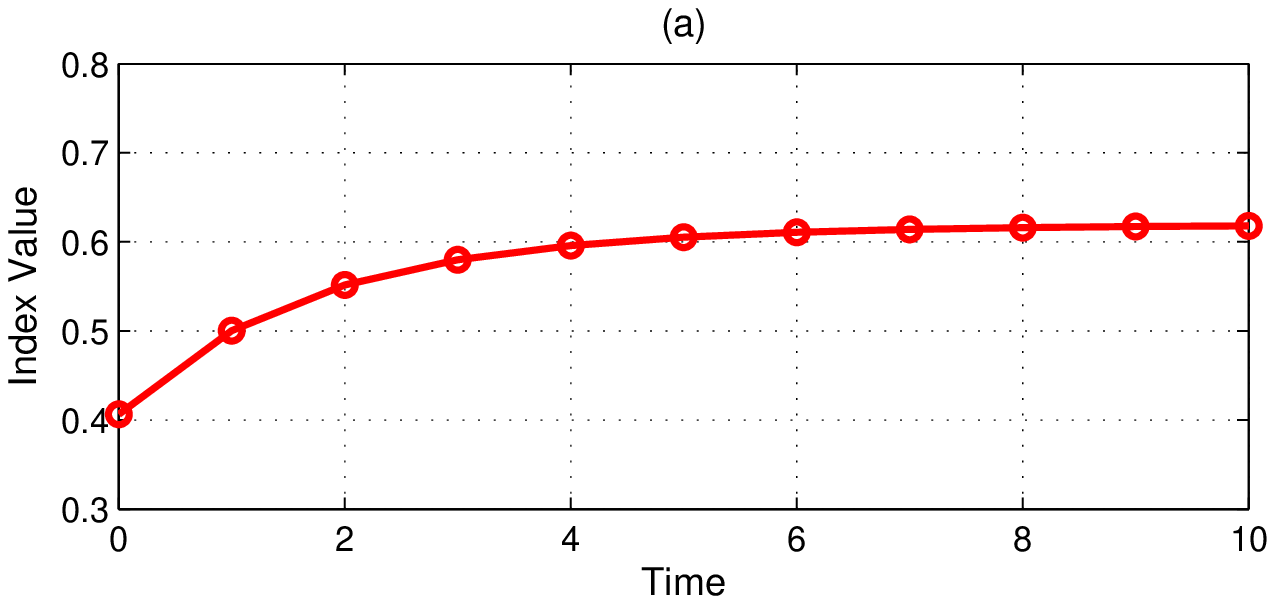}
\includegraphics[width=2.6in]{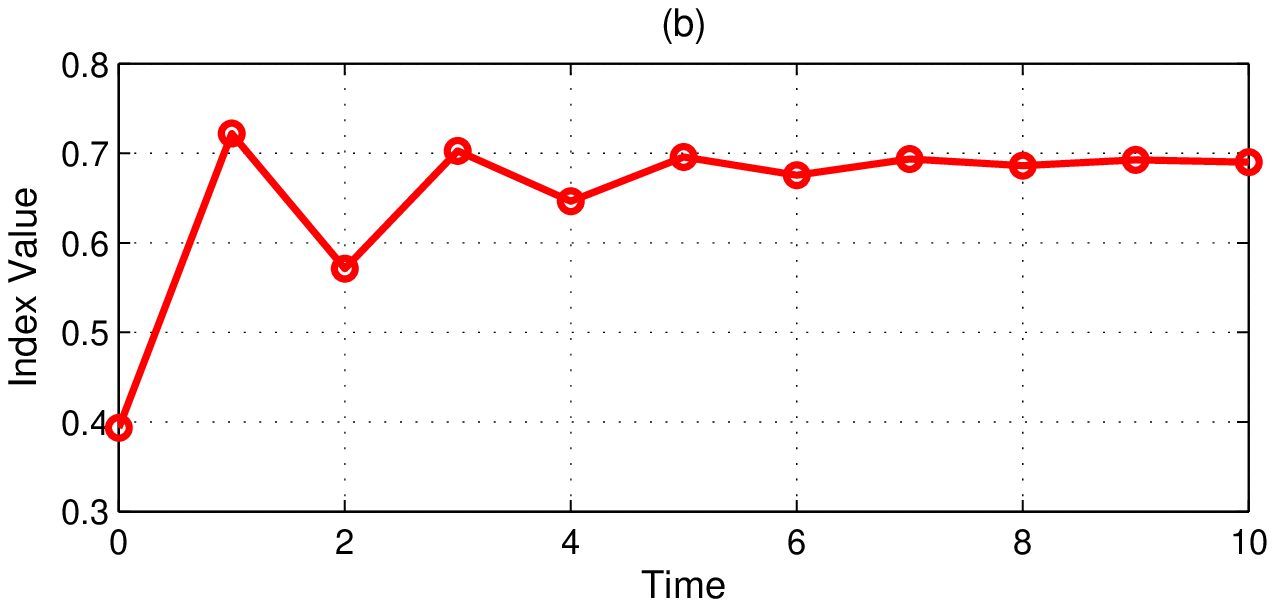}
\vspace{-7.5pt}
\caption{Index value evolution of user $i$, with $\pi_i[0]=0.3$. (a) Positive correlation, $p_i{=}0.8,r_i{=}0.2$; (b) Negative correlation, $p_i{=}0.2,r_i{=}0.8$. }
\label{fig:index}
\vspace{-8pt}
\end{figure}



Note that, from the definition of indexability, the index value $W_i(\pi_i)$ monotonically increases with $\pi_i$. Therefore, when the Markovian channels have the same Markovian structure and vary independently across users (hence the state-index mappings are the same across users), Whittle's index policy essentially becomes the \textit{greedy} policy -- schedule the user with the highest belief value.

Fig.~\ref{fig:index} plots an example of the index value evolution for the case of positively correlated and negatively correlated channels when they stay idle, i.e., not scheduled for transmission. We see that, for the positively correlated channel, the index value behaves monotonically, while, for the negatively correlated channel, the index value shows oscillation. This resembles the evolution of
the belief values, which, as proven in Lemma~\ref{lemma:monotone} in Appendix~\ref{appen:VpVr}, approaches steady state monotonically for the positively correlated channel, and with oscillation for the negatively correlated channel. This resemblance in Fig.~\ref{fig:index} is expected since, from Proposition~\ref{prop:Indexability}, we can infer that the index value monotonically increases with the belief value. Thus, in essence, from Proposition~\ref{prop:Indexability} and Fig.~\ref{fig:index}, we see that the index value captures the underlying dynamics of the Markovian channel.



\begin{figure}
\centering
\includegraphics[width=2.8in]{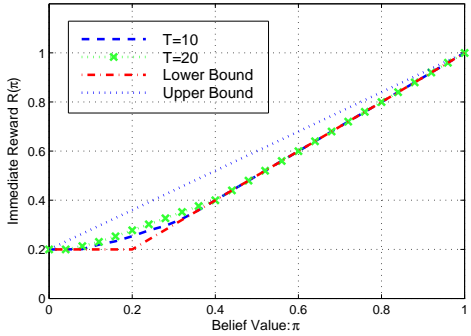}
\vspace{-11pt}
\caption{Immediate reward versus $\pi$.} \label{fig:RwithPi}
\vspace{-4pt}
\end{figure}

\section{Numerical Performance Analysis}
\label{sec:numerical}
\subsection{Model for Simulation}
In this section, we study, via numerical evaluations, the performance of Whittle's index policy, henceforth simply the index policy, for joint estimation and scheduling in our downlink system. We consider the specific class of estimator and rate adapter structure, with pilot-aided training, discussed in Section~\ref{sec:SysModel} and illustrated in Fig.~\ref{fig:block}. We consider a fading channel with the fading coefficients quantized into two levels to reflect the two states of the Markov chain. Additive noise is assumed to be white Gaussian. The channel input-output model is given by $Y= h X+ \epsilon$, where $X, Y$ correspond to transmitted and received signals, respectively, $h$ is the complex fading coefficient and $\epsilon$ is the complex Gaussian, unit variance additive noise. Conditioned on $h$, the Shannon capacity of the channel is given by $R=\log (1+ |h|^2)$. We quantize the fading coefficients such that the allowed rate at the lower state, $\delta=0.2$ for all users. The channel state, represented by the fading coefficient, evolves as Markov chain with fading block length $T$.

We consider a class of Linear Minimum Mean Square Error (LMMSE) estimators \cite{Hassibi2} denoted as $\Phi$. LMMSE estimators are attractive because with additive white Gaussian noise, they can be characterized in closed form \cite{Hassibi2} and, hence, can be conveniently used in simulation. Let $\phi_{\pi}$ denote the optimal LMMSE estimator with prior $\{\pi,1-\pi\}$. We let $\Phi$ denote the set of LMMSE estimators optimized for various values of $\pi$.


\subsection{Immediate Reward Structure}

We now study the structure of the immediate reward $R(\pi)$. Note that $R(\pi)$ is optimized over the class of estimators $\Phi$. Fig.~\ref{fig:RwithPi} illustrates $R(\pi)$, in comparison with the upper and lower bounds derived in Lemma~\ref{lemma:VpVr}, for two values of block length $T$. As established in Lemma~\ref{lemma:VpVr}, $R(\pi)$ shows a convex increasing structure and takes values within the bounds. Note that $R(\pi)$ also increases with $T$, since a larger $T$ provides more channel uses for channel probing and data transmission.

\subsection{Near-optimal Performance of Whittle's Index Policy}


\begin{figure}
\centering
\includegraphics[width=2.9in]{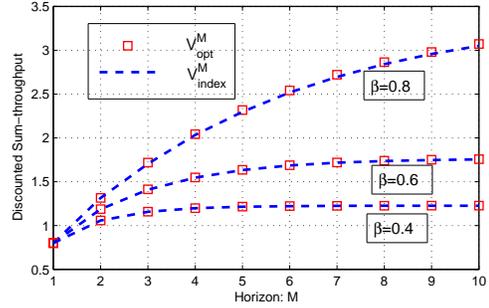}
\vspace{-6pt}
\caption{Performance of the index policy in comparison with that of the optimal policy. System parameters used: N=5, $\{p_1{=}0.2,r_1{=}0.75\}$, $\{p_2=0.6,r_2=0.25\}$, $\{p_3=0.8,r_3=0.3\}$, $\{p_4=0.4,r_4{=}0.7\}$, $\{p_5=0.65,r_5=0.55\}$; Fading block length: $T=20$.}
\label{fig:Vwitht}
\vspace{-10pt}
\end{figure}

We proceed to evaluate the performance of the index policy and compare it with the optimal policy. In Fig.~\ref{fig:Vwitht}, we compare the expected rewards $V^M_{opt}$  and $V^M_{index}$ that, respectively,
correspond to the optimal finite M-horizon policy and the
index policy, for increasing horizon length $M$ and randomly generated system parameters. The value of $V^M_{opt}$ is obtained via brute-force search over the finite horizon. Fig.~\ref{fig:Vwitht} illustrates the near optimal performance of the index policy. Also, as expected, the higher the value of $\beta$, the higher the expected reward.

Table~\ref{tab:tab1} presents the performance of the index policy in a larger perspective. Here, with randomly generated system parameters, the infinite horizon reward under the index policy is compared with those of the optimal policy and a policy that `throws away' the feedback from the scheduled user. Let $V_{no fb}$ denote the reward under this `no feedback' policy. The infinite horizon rewards are obtained as limits of the finite horizon until $1\%$  convergence is achieved. The high values of the quantity $\%$gain${=}\frac{V_{index}{-}V_{no fb}}{V_{opt}-V_{no fb}}{\times} 100\%$, in addition to underscoring the near-optimality of the index policy, also signifies the high system-level gains from exploiting the channel memory using the end-of-slot feedback.

\begin{table}
\vspace{3pt}
\begin{center}
\renewcommand{\tabcolsep}{.3cm}
\renewcommand{\arraystretch}{1.2}
\begin{tabular}{|c|c|c|c|c|c|c|c|c|}
\hline
$N$ & $\beta$ & $V_{opt}$ & $V_{index}$ & $V_{no fb}$ & $\%$gain\\
\hline
4  &  0.6337 &  1.6289  &  1.6289  &  1.4887  & 100$~\%$\\
\hline
4  &  0.5896  &  1.5977  &  1.5866  &  1.2888  & 96.4045$~\%$\\
\hline
4  &  0.6673  &  1.6537  &  1.6319  &  1.4342  & 90.0500$~\%$\\
\hline
5  &  0.4537 &  0.9854  &  0.9854  &  0.9299  & 100$~\%$\\
\hline
5  &  0.6082  &  1.6132  &  1.6072  &  1.4777  & 95.5518$~\%$\\
\hline
5  &  0.6537  &  2.3728  &  2.3725  &  2.1494  & 99.8697$~\%$\\
\hline
5  &  0.5397  &  1.6330  &  1.6330  &  1.5961  & 100$~\%$\\
\hline
\end{tabular}
\end{center}
\caption{Illustration of the gains associated with exploiting channel momory.}
\label{tab:tab1}
\vspace{-14pt}
\end{table}

\begin{figure}
\centering
\includegraphics[width=2.8in]{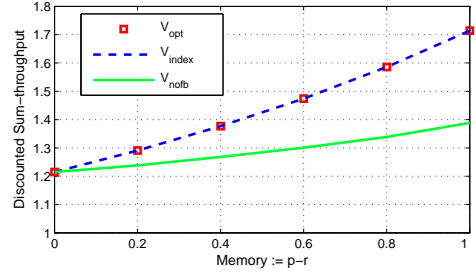}
\vspace{-8pt}
\caption{Illustration of the influence of channel `memory', $(p-r)$, on the performance of the index policy and baseline policies, when $\beta=0.6$.}
\label{fig:VwithMem}
\end{figure}

In Fig.~\ref{fig:VwithMem} we study the effect of the channel `memory' on the performances of
various baseline policies. We consider five users with statistically identical but independently
varying channels. Thus $p_i=p$, $r_i=r$, $i\in\{1, \cdots, 5\}$.  We define the channel `memory' as the difference
$p-r$ and increase the memory by increasing $p$ from $0.5$ to $1$ and maintaining $r=1-p$.
Note that, with this approach, $p+r=1$. Under this condition, the steady state probability that a
channel is in the higher state $h$ is kept constant under varying channel memory. This, essentially,
provides a degree of fairness between systems with different channel memories.  Fig.~\ref{fig:VwithMem} compares the rewards $V_{opt}$, $V_{index}$ and $V_{nofb}$  that respectively correspond to the rewards under the optimal policy, the index policy, and the `no feedback' policy introduced earlier, for increasing channel memory. Note that when $p=r$, the channel of each user evolves \emph{i.i.d.} across time, with no information contained in the channel state feedback. Thus the policy that throws away this feedback achieves the same performance as the optimal policy that optimally uses this feedback, i.e., $V_{nofb}=V_{opt}$ when $p=r$. Also, since the channels are \emph{i.i.d.} across users, when $p=r$, the index policy simplifies to a `randomized' policy that schedules randomly and uniformly across users, in effect mirroring the `no feedback' policy in this setting. This explains $V_{index}=V_{nofb}$ when $p=r$. As the channel memory increases, the significance of the channel state feedback increases, resulting in an increasing gap between the policies that use this feedback (optimal and Whittle's index policies) and the `no feedback' policy.

Fig.~\ref{fig:VwithMem}, along with Table~\ref{tab:tab1}, shows that exploiting channel memory for opportunistic scheduling can result in significant performance gains, and almost all of these gains can be realized using the easy-to-implement index policy.
\vspace{-3pt}

\section{Conclusion}

In this paper, we have studied downlink multiuser scheduling under a Markov-modeled channel. We considered the scenario in which the channel state information is not perfectly known at the scheduler, essentially requiring a joint design of user selection, channel estimation and rate adaptation. This calls for a two-stage optimization: (1) Within each slot, the channel estimation and rate adaptation is optimized to obtain an optimal transmission rate in the scheduling slot; (2) Across scheduling slots, users are selected to maximize the infinite horizon discounted reward. We formulated the scheduling problem as a partially observable Markov decision processes with the classic `exploitation versus exploration' trade-off. We then linked the problem to a restless multiarmed bandit processes and conducted a Whittle's indexability analysis. By obtaining structural properties of the optimal reward within the indexability setup, we showed that the downlink scheduling problem is Whittle indexable. We then explicitly characterized the Whittle's index policy and studied the performance of this policy using extensive numerical experiments, which suggest that the index policy has near optimal performance and significant system level gains can be realized by exploiting the channel memory for joint channel estimation and scheduling.
\vspace{4pt}

\appendices

\section{Proof of Lemma~\ref{lemma:VpVr}}
\label{appen:VpVr}

We first establish structural properties of the belief update when a user stays idle. Suppose a user has the initial belief value $\pi[0]$ and stays idle at all times, the belief value at $t^{th}$ slot is then given by $\pi[t]=Q^t(\pi_i[0])$, where $Q^t$ is the $t^{th}$ iteration of function $Q$, given by
\begin{align}
\label{eq:Qupdate}
Q^t(\pi)= \frac{r-(p-r)^t\big(r-(1+r-p)\pi\big)}{1+r-p}.
\end{align}

We let $\pi^0$ be the steady state distribution of the two-state channel being at the higher state, i.e.,
\begin{align}
\nonumber
\pi^0=\frac{r}{1+r-p}. \nonumber
\end{align}

\begin{figure}
\centering
\includegraphics[width=2.8in]{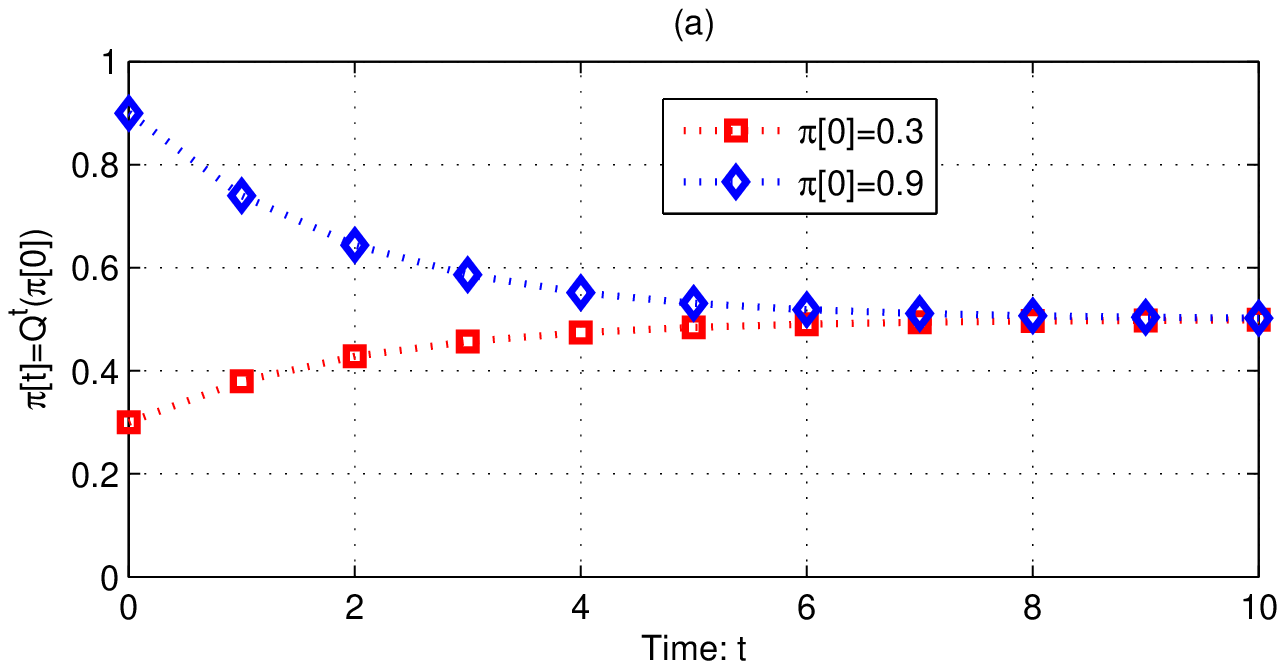}
\includegraphics[width=2.8in]{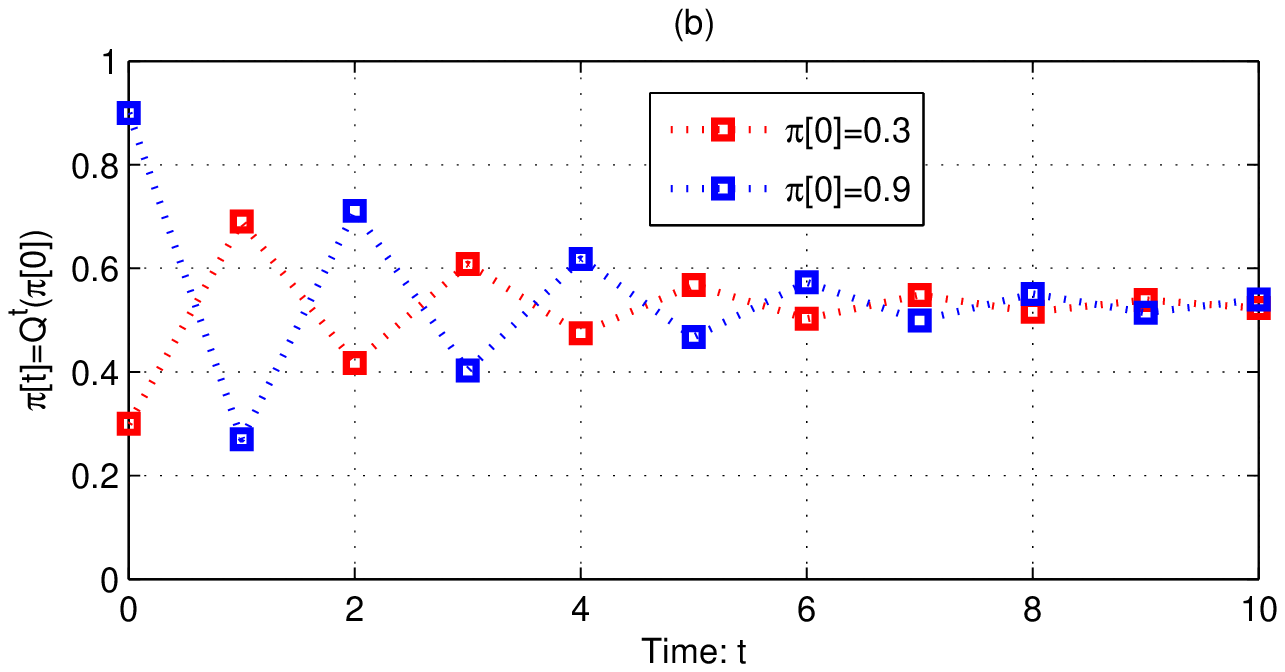}
\caption{Evolution of belief values under consecutive idle decisions. (a) Positive correlation, $p=0.8$, $r=0.2$; (b) Negative correlation, $p=0.2$, $r=0.8$.}
\label{fig:Qupdate}
\end{figure}

It is clear that $\pi^0=\lim_{t\rightarrow \infty}Q^t(\pi).$ An example of the  belief evolution when a user stays idle is depicted in Fig.~\ref{fig:Qupdate}. This figure shows that, when staying idle, the belief value approaches steady state monotonically for positively correlated channel and approaches steady state with oscillation for negatively correlated channel. The structural properties of $Q^t(\pi_i[0])$ is critical to the rest of the proof and is recorded in the following lemma.

\begin{lemma}\label{lemma:monotone}
\quad\\
(i) For positively correlated channel (i.e., $p>r$), $\pi[t]$ converges to steady state $\pi^0$ monotonically. For negatively correlated channel (i.e., $p\leq r$), $\pi[t]$ converges to steady state $\pi^0$ with oscillation and a monotonically converging envelope.

\noindent(ii) $\min\{p,r\} \leq Q^t(\pi_i[0]) \leq \max\{p,r\}$ for all $t=1,2, \cdots$ and $\pi_i[0] \in [0,1]$.

\end{lemma}

\newcounter{TempEqCnt2}
\setcounter{TempEqCnt2}{\value{equation}}
\setcounter{equation}{13}

\begin{figure*}[hb]
\hrulefill
\begin{align}
\label{eq:Vomega}
V_{\omega}(r)=\frac{(1{-}\beta^{L(r,\pi^*(\omega))})(1-\beta p)\omega{+}(1{-}\beta)
\beta^{L(r, \pi^*(\omega))}\big[(1-\beta p)R(Q^{L(r,\pi^*(\omega))}(r)){+}\beta
Q^{L(r,\pi^*(\omega))}(r)R(p) \big]}{(1-\beta)(1-\beta p)\big(1-\beta^{L(r,\pi^*(\omega))+1}\big)+(1-\beta)^2
Q^{L(r,\pi^*(\omega))}(r)\beta^{L(r, \pi^*(\omega))}}
\end{align}
\vspace{-0.2in}
\end{figure*}

\setcounter{equation}{\value{TempEqCnt2}}

\noindent{\textbf{Proof:}}
(i) Since we have $0< p-r \leq 1$ for positively correlated channel and $-1\leq p-r \leq 0$ for negatively correlated channel, it is clear from the expression of~(\ref{eq:Qupdate}) that $\pi[t]$ converges to steady state $\pi^0$ monotonically and approaches steady state $\pi^0$ with oscillation and a monotonically converging envelop.

\noindent(ii) Since we have established part (i), it suffices to check that the first step transition satisfies: $\min\{p,r\} \leq Q[\pi]\leq \max\{p,r\}$, for all $\pi$, as shown below.
\begin{align}
Q(\pi)= \frac{r-(p-r)\big(r-(1+r-p)\pi\big)}{1+r-p}. \nonumber
\end{align}

For positively correlated channel, since $p-r>0$
\begin{align}
Q(\pi)\geq& \frac{r-(p-r)r}{1+r-p}=r.\nonumber\\
Q(\pi)\leq& \frac{r{-}(p{-}r)\big(r{-}(1{+}r{-}p)\big)}{1+r-p}=\frac{p(1-p+r)}{1+r-p}=p. \nonumber
\end{align}

For negatively correlated channel, since $p-r\leq 0$,
\begin{align}
Q(\pi)\leq& \frac{r-(p-r)r}{1+r-p}=r.\nonumber\\
Q(\pi)\geq& \frac{r{-}(p{-}r)\big(r{-}(1{+}r{-}p)\big)}{1+r-p}=\frac{p(1{-}p{+}r)}{1+r-p}=p. \nonumber
\end{align}

The lemma is thus proved. $\hfill \blacksquare$
\vspace{5pt}

We then define $L(\pi, \pi^*)$ as the time needed for belief value of a user to exceed
$\pi^*$ from below, starting from initial value $\pi$. Formally,
\begin{align}
L(\pi, \pi^*)=\min_t \{Q^t(\pi)> \pi^* \} \nonumber
\end{align}

Using Lemma~\ref{lemma:monotone} and expression~(\ref{eq:Qupdate}), $L(\pi, \pi^*)$ can be calculated as follows.
\vspace{8pt}

$\bullet$ Positive correlation ($p>r$)

$L(\pi, \pi^*)=
\begin{cases}
0 &\text{if $\pi > \pi^*$}\\
\lfloor log_{p{-}r} \frac{r-(1+r-p)\pi^*}{r-(1+r-p)\pi} \rfloor {+}1  &\text{if $\pi {\leq} \pi^* {<} \pi_0$}\\
\infty \hspace{1in} \text{if $\pi {\leq} \pi^*$}&\text{and }  \text{$\pi^*{\geq} \pi_0$}\\
\end{cases}$
\vspace{7pt}

$\bullet$ Negative correlation ($p\leq r$)

$L(\pi, \pi^*)=
\begin{cases}
0 &\text{ if $\pi > \pi^*;$}\\
1 &\text{ if
$\pi \leq \pi^*$ and $ Q(\pi) > \pi^*$,}\\
\infty &\text{ if $\pi \leq \pi^*$ and  $Q(\pi) \leq \pi^*.$}\\
\end{cases}$
\vspace{8pt}

We shall refer to the `active set' as the set of belief values for which the optimal decision is to transmit. The `idle set' denotes the set of belief values for which the optimal decision is to stay idle. We proceed to derive the value functions $V_{\omega}(p)$ and $V_{\omega}(r)$ based on the value of $\pi^*(\omega)$.
\vspace{8pt}

(1) Positive correlation ($p>r$).
\vspace{8pt}

$\bullet$ When $\pi^*(\omega) \geq p$, the belief value $p$ is thus in the `idle set'. From Lemma~\ref{lemma:monotone}(ii), if $\pi[0]=p$, the system stays idle. Hence the reward function is expressed as
\begin{align}
V_{\omega}(p)=\omega+\beta \omega+ \beta^2 \omega+ \cdots =\frac{\omega}{1-\beta}. \nonumber
\end{align}

$\bullet$ When $\pi^*(\omega) < p$, the belief value $p$ is then in the `active set'. Hence from the Bellman equation in~(\ref{eq:Bellman}),
\begin{align}
V_{\omega}(p)=R(p)+\beta \big(pV_{\omega}(p)+(1-p)V_{\omega}(r)\big). \nonumber
\end{align}

Rearranging the terms yields,
\begin{align}
V_{\omega}(p)=\frac{R(p)+\beta(1-p)V_{\omega}(r)}{1-\beta p}. \nonumber
\end{align}

$\bullet$ When $\pi^*(\omega) < r$, the value $r$ is then in `active set'. From Lemma~\ref{lemma:monotone}(ii), regardless of the scheduling decision, the belief values $\pi[t]$, starting from $\pi[0]=r$, stays in the `active set'. Therefore
\begin{align}
V_{\omega}(r)&=\sum_{t=0}^{\infty} \beta^t R(Q^{t}(r)) =\sum_{t=0}^{\infty} \beta^t R\Big(\frac{r{-}(p{-}r)^{t{+}1}r}{1{+}r{-}p}\Big).\nonumber
\end{align}

$\bullet$ When $\pi^*(\omega) \geq \pi^0$, since $\pi^0 \geq r$, the belief value $r$ is in `idle set'. From Lemma \ref{lemma:monotone}(i), the belief values $\pi[t]$, starting from $\pi[0]=r$, stays in `idle set'. Hence
\begin{align}
\nonumber
V_{\omega}(r)=\omega+\beta \omega+ \beta^2 \omega+ \cdots =\frac{\omega}{1-\beta}.
\end{align}

$\bullet$ When $r < \pi^*(\omega) < \pi^{0}$, the belief value $r$ is therefore in `idle set'. Since the channel is positively correlated, from Lemma~\ref{lemma:monotone}, starting from $\pi[0]=r$, the user remains idle for a duration of $L(r,\pi^*(\omega))$ slots. Therefore
\begin{align}
\hspace{-11pt}V_{\omega}(r){=}\frac{1{-}\beta^{L(r,\pi^*(\omega))}}{1-\beta} \omega{+}\beta^{L(r,\pi^*(\omega))}V_{\omega}^1\big(Q^{L(r,\pi^*(\omega))}(r)\big).\label{eq:Vr1}
\end{align}
where
\begin{align}
&V^1_{\omega}\big(Q^{L(r,\pi^*(\omega))}(r)\big)=R\big(Q^{L(r,\pi^*(\omega))}(r)\big)+\nonumber \\
 &\hspace{27pt}\beta \big(Q^{L(r,\pi^*(\omega))}(r)V_{\omega}(p)+ (1-Q^{L(r,\pi^*(\omega))}(r))V_{\omega}(r)\big) \nonumber
\end{align}

Substituting the above expression in~(\ref{eq:Vr1}), we can obtain $V_{\omega}(r)=\Theta$ as given in expression~(\ref{eq: Theta}) in the lemma.

\vspace{8pt}

(2) Negative correlation ($p\leq r$).
\vspace{2pt}

The derivation of $V_{\omega}(p)$ and $V_{\omega}(r)$ for negative correlation case follows an approach similar to that for the case of positive correlation. Details are, therefore, omitted here. $\hfill \blacksquare$
\vspace{7pt}

Note that the expressions of the value functions $V_{\omega}(p)$ and $V_{\omega}(r)$ in Lemma~\ref{lemma:VpVr} are not in closed form. However, the closed form expressions for $V_{\omega}(p)$ and $V_{\omega}(r)$ can be easily calculated based on the expressions in Lemma~\ref{lemma:VpVr}, recorded below.
\vspace{5pt}

\noindent Case (1) Positive correlation ($p > r$). First we give the closed form expression of $V_{\omega}(p)$.
\vspace{4pt}

$\bullet$ If $\pi^*(\omega) < \pi^0$,
\begin{align}
V_{\omega}(p)=\sum_{t{=}0}^{\infty} \beta^t
R\Big(\frac{r{+}(p{-}r)^{t{+}1}(1{-}p)}{1{+}r{-}p}\Big). \nonumber
\end{align}

$\bullet$ If $\pi^0 \leq\pi^*(\omega) < p$,
\begin{align}
V_{\omega}(p)=\frac{\beta(1-p)\omega+(1-\beta)R(p)}{(1-\beta)(1-\beta p)}. \nonumber
\end{align}

$\bullet$ If $\pi^*(\omega) \geq p$, $V_{\omega}(p)=\omega /(1-\beta).$
\vspace{8pt}

We proceed to give the closed form expression of $V_{\omega}(r)$.
\vspace{3pt}

$\bullet$ If $\pi^*(\omega) < r$,
\begin{align}
V_{\omega}(r)=\sum_{t=0}^{\infty} \beta^t R\Big(\frac{r-(p-r)^{t+1}r}{1+r-p}\Big).\nonumber
\end{align}

$\bullet$ If $r {\leq} \pi^*(\omega) {<} \pi^0$, $V_{\omega}(r)$ is given in equation~(\ref{eq:Vomega}).
\setcounter{equation}{14}

$\bullet$ If $\pi^*(\omega) \geq \pi^0$, $V_{\omega}(r)=\omega/(1-\beta)$.

\vspace{8pt}

\noindent Case (2) Negative correlation ($p \leq r$). In this case, the closed form expression of $V_{\omega}(r)$ is given as follows.
\vspace{4pt}

$\bullet$ If $\pi^*(\omega) \geq  r$, then $V_{\omega}(r)= \omega/(1-\beta). $

$\bullet$ If $Q(p) \leq \pi^*(\omega) < r$, then
\begin{align}
V_{\omega}(r)=\frac{\beta r \omega +(1-\beta)R(r)}{(1-\beta)(1-\beta(1-r))} \nonumber.
\end{align}

$\bullet$ If $p \leq \pi^*(\omega) < Q(p)$, we have
\begin{align}
V_{\omega}(r)=\frac{\beta r \omega + \beta^2 r R(Q(p))+(1-\beta^2 Q(p))R(r)}{(1-\beta(1-r))(1{-}\beta^2 Q(p))-\beta^3 r (1-Q(p))} \nonumber .
\end{align}

$\bullet$ If $\pi^*(\omega) < p$, then
\begin{align}
V_{\omega}(r)=\sum_{t=0}^{\infty} \beta^t R\Big(\frac{r-(p-r)^{t+1}r}{1+r-p}\Big) \nonumber.
\end{align}

Then we give the closed form expression of $V_{\omega}(p)$.
\vspace{7pt}

$\bullet$ If $\pi^*(\omega) <  p$, then
\begin{align}
V_{\omega}(p)=\sum_{t=0}^{\infty} \beta^t
R\Big(\frac{r+(p-r)^{t+1}(1-p)}{1+r-p}\Big). \nonumber
\end{align}

$\bullet$ If $p {\leq} \pi^*(\omega) < Q(p)$, we have
\begin{align}
V_{\omega}(p){=}\frac{\big(1{-}\beta(1{-}r)\big)\big[\omega{+}\beta R\big(Q(p)\big)\big]{+}\beta^2 \big(1{-}Q(p)\big) R(r)}{\big(1-\beta(1-r)\big)\big(1-\beta^2 Q(p)\big)-\beta^3 r \big(1-Q(p)\big)}. \nonumber
\end{align}

$\bullet$ If $\pi^*(\omega) {\geq} Q(p)$, then $V_{\omega}(p)=\omega / (1-\beta).$

\section{Proof of Proposition~\ref{prop:Indexability}}
\label{appen:Indexability}

We prove that the problem is Whittle indexable by showing that $\pi^*(\omega)$ monotonically increases with $\omega$. It is clear from Proposition~\ref{prop:thresd} that $\pi^*(\omega)=\kappa$ for $\omega \in [0,\delta)$. So it suffices to show that $\pi^*(\omega)$ is strictly increasing for $\omega \in [\delta,1]$. The proof technique follows along the lines of \cite{Zhao_index} and is presented next. We first proceed with the following lemma, where the right derivative of the reward function is compared.

\begin{lemma}
If for all $\omega \in [\delta,1]$, we have
\begin{align}
\label{eq:obj}
\frac{d V^1_{\omega}(\pi)}{d \omega} \big |_{\pi=\pi^*(\omega)} < \frac{d V^0_{\omega}(\pi)}{d \omega} \big |_{\pi=\pi^*(\omega)},
\end{align}
then ${\pi^*(\omega)}$ is strictly increasing with $\omega$ for $\omega \in [\delta,1]$.
\end{lemma}

\noindent\textbf{Proof:}
The lemma is proven by contradiction. Suppose there exists $\omega_0 \in [\delta, 1]$, such that $\pi^*(\omega)$ is decreasing (i.e., non-increasing) at $\omega_0$, hence it is decreasing in a neighborhood of $\omega_0$, say, $[\omega_0, \omega_0+\triangle \omega]$. Since $V^1_{\omega_0+\triangle \omega}\big(\pi^*(\omega_0+\triangle \omega)\big)=V^0_{\omega_0+\triangle \omega}\big(\pi^*(\omega_0+\triangle \omega)\big)$ and $\pi^*(\omega)$ is decreasing at $\omega_0$, $\pi^*(\omega_0)$ is within the `active set' for the $(\omega_0+\triangle \omega)$-subsidy problem. Therefore we have $V^1_{\omega_0+\triangle \omega}\big(\pi^*(\omega_0)\big)\geq V^0_{\omega_0+\triangle \omega}\big(\pi^*(\omega_0)\big)$. Besides, from the definition of threshold value $\pi^*(\omega_0)$, $V^1_{\omega_0}\big(\pi^*(\omega_0)\big)= V^0_{\omega_0}\big(\pi^*(\omega_0)\big)$. Therefore,
\begin{align}
\frac{d V^1_{\omega}(\pi)}{d \omega} \big |_{\pi=\pi^*(\omega)}&=\lim_{\triangle \omega \rightarrow 0}\frac{V^1_{\omega_0+\triangle \omega}\big(\pi^*(\omega_0)\big){-}V^1_{\omega_0}\big(\pi^*(\omega_0)\big)}{\triangle \omega} \nonumber \\
&\geq \lim_{\triangle \omega \rightarrow 0} \frac{V^0_{\omega_0+\triangle \omega}\big(\pi^*(\omega_0)\big){-}V^0_{\omega_0}\big(\pi^*(\omega_0)\big)}{\triangle \omega} \nonumber \\
&=\frac{d V^0_{\omega}(\pi)}{d \omega} \big
|_{\pi=\pi^*(\omega)}, \nonumber
\end{align}
which contradicts with the assumption. $\hfill \blacksquare$

\vspace{7pt}

Therefore, to establish indexability, it suffices to prove the inequality~(\ref{eq:obj}), i.e., $\frac{d V^1_{\omega}(\pi)}{d \omega} \big |_{\pi=\pi^*(\omega)} < \frac{d V^0_{\omega}(\pi)}{d \omega} \big |_{\pi=\pi^*(\omega)}$. Let $D_{\omega}(\pi)$ be the
discounted time the $\omega$-subsidy process, with initial belief $\pi$, is made passive, i.e.,
\begin{align}\nonumber
D_{\omega}(\pi)=\sum_{t=0}^{\infty} \beta^t \mathbb{1} (a[t]=0).
\end{align}

Noting that giving the value of $\pi^*(\omega)$, the studying the belief value evolution follows the same pattern as in ON/OFF channel case, hence the expression of $D_{\omega}(\pi)$ takes the same form as given in \cite{Zhao_index}. It follows from \cite{Whittle}\cite{Zhao_index} that $D_{\omega}(\pi)=\frac{d V_{\omega}(\pi)}{d \omega}$. Taking derivative of the $V^1_{\omega}(\pi)$ and $V^2_{\omega}(\pi)$ expressions~(\ref{eq:V1_Bellman})-(\ref{eq:V0_Bellman}) with respect to $\omega$, the objective (\ref{eq:obj}) now becomes
\begin{align}
\label{eq:obj2}
\beta\big(\pi^*(\omega) D_{\omega}(p){+}(1{-}\pi^*(\omega))D_{\omega}(r)\big) {<}1{+} \beta D_{\omega}\big(Q(\pi^*(\omega))\big).
\end{align}

\noindent Case (1) If $0 \leq \pi^*(\omega) < \min \{ p, r \}$, from Lemma~\ref{lemma:monotone}(ii), starting from the initial belief value $\pi[0]=r$ or $\pi[0]=p$, the believe value $\pi[t]$ never evolves below $\pi^*(\omega)$, hence the project is active at all times under optimal control. Therefore $D_{\omega}(p)=D_{\omega}(r)=D_{\omega}(Q(\pi^*(\omega)))=0$. Equation (\ref{eq:obj2}) thus holds.
\vspace{8pt}

\newcounter{TempEqCnt3}
\setcounter{TempEqCnt3}{\value{equation}}
\setcounter{equation}{20}
\begin{figure*}[hb]
\hrulefill
\vspace{5pt}
\begin{align}
\label{eq:W1}
\hspace{-10pt}W_i(\pi_i)=\frac{\big[R_i(\pi_i){-}\beta R_i(Q_i(\pi_i)){+}\beta\big (\pi_i{-}\beta Q_i(\pi_i)\big)\sum_{t{=}0}^{\infty} \beta^t R(\frac{r_i{+}(p_i{-}r_i)^{t{+}1}(1{-}p_i)}{1{+}r_i{-}p_i})\big] \Gamma_i+\beta\big[(1{-}\pi_i){-}\beta(1{-}Q_i(\pi_i))\big]\Lambda_i }{\Gamma_i-\beta(1{-}\beta^{L(r_i,\pi)})(1-\beta p)\big[(1-\pi_i)-\beta \big(1-Q_i(\pi_i)\big)\big]}
\end{align}

\begin{align}
W_i(\pi_i)=&[R_i(\pi_i)-\beta R_i(Q_i(\pi_i))]+\beta \big(\pi_i-\beta Q_i(\pi_i)\big)\cdot \sum_{t=0}^{\infty} \beta^t R\Big(\frac{r_i+(p_i{-}r_i)^{t+1}(1-p_i)}{1+r_i{-}p_i}\Big)+\nonumber\\
&\hspace{0.7in}\beta[(1-\pi_i)-\beta(1-Q_i(\pi_i))]\cdot \sum_{t=0}^{\infty} \beta^t R\Big(\frac{r-(p-r)^{t+1}r}{1+r-p}\Big). \label{eq:W2}
\end{align}
\setcounter{equation}{25}

\begin{align}
W_i(\pi_i){=}\frac{\big[R_i(\pi_i)-\beta R_i(Q_i(\pi_i))\big]\cdot \Delta_i+\beta[\pi_i-\beta Q_i(\pi_i)] \cdot \Omega_i + \beta [(1-\pi_i)-\beta (1-Q_i(\pi_i))] \cdot \Upsilon_i}{\Delta_i -\beta(\pi_i-\beta Q_i(\pi_i))(1-\beta(1-r_i))-\beta^2 r_i[(1-\pi_i)-\beta (1-Q_i(\pi_i))]} \label{eq:W3}
\end{align}

\begin{align}
W_i(\pi_i)=&\big[R(\pi_i)-\beta R_i(Q_i(\pi_i))\big]+ \beta \big[\pi_i-\beta Q_i(\pi_i)\big] \sum_{t=0}^{\infty} \beta^t R\Big(\frac{r_i+(p_i-r_i)^{t+1}(1-p_i)}{1+r_i-p_i}\Big)+\nonumber\\ &\hspace{0.5in}\beta\big[(1-\pi_i)-\beta\big(1-Q_i(\pi_i)\big)]\sum_{t=0}^{\infty} \beta^t R_i\Big(\frac{r_i-(p_i-r_i)^{t+1}r_i}{1+r_i-p_i}\Big). \label{eq:W4}
\end{align}

\vspace{-0.2in}
\end{figure*}

\setcounter{equation}{\value{TempEqCnt3}}

\noindent Case (2) If $\pi_0 \leq \pi^*(\omega) \leq 1$, starting from initial belief $\pi[0]=Q(\pi^*(\omega))$, the belief value $\pi[t]$ always stays within the `idle set', i.e., $D_{\omega}(Q(\pi^*(\omega)))=\frac{1}{1-\beta}$. Equation (\ref{eq:obj2}) holds since $D_{\omega}(p)\leq 1+\beta+\beta^2+\cdots = \frac{1}{1-\beta}$ and, similarly, $D_{\omega}(r)\leq \frac{1}{1-\beta}$.
\vspace{1pt}

\noindent Case (3) If $\min \{ p, r \} \leq \pi^*(\omega) \leq \pi_0$, from Lemma~\ref{lemma:monotone}(ii), $Q(\pi^*(\omega))$ is in `active set'. Since
\begin{align}
&V_{\omega}(Q(\pi^*(\omega)))\nonumber \\
=&R(Q(\pi^*(\omega))+ \beta [Q(\pi^*(\omega))
V_{\omega}(p) +(1-Q(\pi^*(\omega))) V_{\omega}(r)], \nonumber
\end{align}
we have
\begin{align}
\label{eq:D_omegaQ}
&D_{\omega}(Q(\pi^*(\omega)))\nonumber \\
=& \beta [Q(\pi^*(\omega))
D_{\omega}(p) +(1-Q(\pi^*(\omega))) D_{\omega}(r)].
\end{align}

We then discuss~(\ref{eq:D_omegaQ}) separately for negatively and positively correlated channels.
\vspace{5pt}

$\bullet$ Negatively correlated channel ($p\leq r$). Since $r > \pi^0 >\pi^*(\omega)$, the belief value $r$ is in the `active set', hence
\begin{align}
V_{\omega}(r)=R(r)+\beta(r V_{\omega}(p)+(1-r)V_{\omega}(r)). \nonumber
\end{align}

Therefore, we have
\begin{align}
\label{eq:D_omegar}
D_{\omega}(r)=\beta(r D_{\omega}(p)+(1-r)D_{\omega}(r)).
\end{align}

Substituting equation~(\ref{eq:D_omegaQ}) and~(\ref{eq:D_omegar}) in~(\ref{eq:obj2}), we get
\begin{align}
\nonumber
\frac{\beta}{1-\beta(1-r)} D_{\omega}(p) (1-\beta)(\beta
r+\pi^*(\omega)-\beta Q(\pi^*(\omega))) < 1.
\end{align}

Following the same technique as in \cite{Zhao_index}, the above inequality can be verified by substituting $\pi^*(\omega)$ by $\pi^0$ and $D_{\omega}(p)$ by $\frac{1}{1-\beta}$.
\vspace{2pt}

$\bullet$ Positively correlated channel ($p>r$). In this case, $p$ is in the `active set', hence
\begin{align}
V_{\omega}(p)=R(p)+\beta\big(pV_{\omega}(p)+(1-p)V_{\omega}(r)\big). \nonumber
\end{align}

Taking derivative with respect to $\omega$ we have,
\begin{align}
\label{eq:D_omegap}
D_{\omega}(p)=\beta\big(pD_{\omega}(p)+(1-p)D_{\omega}(r)\big).
\end{align}

Substituting equations~(\ref{eq:D_omegaQ}) and (\ref{eq:D_omegap}) in (\ref{eq:obj2}), we have
\begin{align}
\beta D_{\omega}(r) (1-\beta) (1-\frac{\pi^*(\omega)-\beta Q(\pi^*(\omega))}{1-\beta p}) <1. \nonumber
\end{align}

We note that the expression of $D_{\omega}(r)$ takes the same form as in \cite{Zhao_index}. By applying the same technique as in \cite{Zhao_index}, it can be checked that the above inequality indeed holds.

Therefore the inequality~(\ref{eq:obj2}) is justified and hence indexability holds. $\hfill \blacksquare$

\section{Proof of Proposition~\ref{prop:Index_val}}
\label{appen:Index_val}

For $\omega$-subsidy problem of user $i$, from indexability, we know that $\pi_i^*(\omega)$ strictly increases from $0$ to $1$ as $\omega$ increases from $\delta_i$ to $1$. Hence the index value, from its definition in~(\ref{eq:Index_val}), is the subsidy value for which the active and idle decisions are equally attractive. We can hence derive index value $W_i(\pi_i)$ by equating $V^1_{i,\omega}(\pi_i)$ and $V^0_{i, \omega}(\pi_i)$ and solve for $\omega$ as a function of $\pi_i$, i.e.,
\begin{align}
&W_i(\pi_i)+\beta V_{i,W_i(\pi_i)}(Q_i(\pi_i))\nonumber \\
=&R(\pi_i)+\beta [\pi_i
V_{i,W_i(\pi_i)}(p_i)+(1-\pi_i)V_{i,W_i(\pi_i)}(r_i)]. \label{eq:Index_Dir}
\end{align}

Note that the expressions of $V_{i,\omega}(p_i)$ and $V_{i,\omega}(r_i)$ have been given by Lemma~\ref{lemma:VpVr}. Substituting in~(\ref{eq:Index_Dir}) the values of $V_{i,\omega}(p_i)$ and $V_{i,\omega}(r_i)$, we obtain the index value expressions, explained in the following.
\vspace{8pt}

Case (1). Positively correlation ($p_i > r_i$).
\vspace{7pt}

$\bullet$ If $\pi_i \geq p_i$, the belief value $Q_i(\pi_i)$, $p_i$, $r_i$ are in the
`idle set' and, starting from initial belief $\pi_i[0]=Q_i(\pi_i)$ or $\pi_i[0]=p_i$, or $\pi_i[0]=r_i$, $\pi_i[t]$ will stay in the `idle set'. Hence
\begin{align}
V_{i,\omega}(Q_i(\pi_i))=V_{i,\omega}(p_i)=V_{i,\omega}(r_i)=\frac{\omega}{1-\beta}. \nonumber
\end{align}

Substituting the above expressions in~(\ref{eq:Index_Dir}) we obtain that $W_i(\pi_i)=R(\pi_i)$.
\vspace{7pt}

$\bullet$ If $\pi^0_i \leq \pi_i < p_i$, then $p_i$ is in `active set', and starting from initial belief $\pi_i[0]=r_i$ or $\pi_i[0]=Q_i(\pi_i)$, $\pi_i[t]$ stays within `idle set' at all times. Hence
\begin{align}
V_{i,\omega}(Q_i(\pi_i))=V_{i,\omega}(r_i)=\frac{\omega}{1-\beta}. \nonumber
\end{align}

Substituting the above expressions and the expression of $V_{i,\omega}(p_i)$ (given in Lemma~\ref{lemma:VpVr}) in~(\ref{eq:Index_Dir}), we get
\begin{align}
W_i(\pi_i)=\frac{\beta \pi_i R(p_i)+(1-\beta p_i)R(\pi_i)}{1+ \beta \pi_i -\beta p_i}. \nonumber
\end{align}

$\bullet$ If $\pi_i < \pi_i^0 $, then the value $Q_i(\pi_i)$ is in the `active set'. Therefore,
\begin{align}
\hspace{7pt}V_{i,\omega}\big(Q_i(\pi_i)\big)= &R(Q_i(\pi_i))+ \nonumber \\
&\beta[Q_i(\pi_i) V_{i,\omega}(p_i)+
(1-Q_i(\pi_i)) V_{i,\omega}(r_i) ]. \nonumber
\end{align}

Again, substituting the expression of $V_{i,\omega}\big(Q_i(\pi_i)\big)$ in~(\ref{eq:Index_Dir}), we have
\begin{align}
\hspace{7pt}W_i(\pi_i)= &[R(\pi_i){-}\beta R(Q_i(\pi_i))]{+}\beta[ \pi_i {-}\beta
Q_i(\pi_i)]V_{i,W_i(\pi_i)}(p_i)\nonumber \\
&\hspace{17pt}{+}\beta [(1-\pi_i){-}\beta
(1{-}Q_i(\pi_i))]V_{i,W_i(\pi_i)}(r_i). \nonumber
\end{align}

Case (2). Negative correlation ($r_i \geq p_i$).

Using the similar approach as in the positive correlation case, the expressions of the index value can be derived for the case of negative correlation, which are given in the Proposition~\ref{prop:Index_val}. Details are, therefore, omitted here. $\hfill \blacksquare$
\vspace{7pt}

Note that the expressions given in Proposition~\ref{prop:Index_val} are not in closed form. However, the closed form expression for the index value $W_i(\pi_i)$ can be easily calculated based on these expressions provided in Proposition~\ref{prop:Index_val}, recorded as follows.
\vspace{8pt}

\noindent Case (1). Positively correlated channel ($p_i > r_i$).\\

$\bullet$ If $\pi_i \geq p_i$, then the index value $W_i(\pi_i)=R_i(\pi_i)$.

$\bullet$ If $\pi_i^0 \leq \pi_i < p_i$, then
\begin{align}
W_i(\pi_i)=\frac{\beta \pi_i R_i(p_i) + (1-\beta p_i) R_i(\pi_i)
}{1+\beta \pi_i-\beta p_i} \nonumber
\end{align}

$\bullet$ If $r_i \leq \pi_i < \pi_i^0$, $W_i(\pi_i)$ is given in~(\ref{eq:W1}), where
\begin{align}
\Gamma_i&=(1-\beta)(1-\beta p)\big(1-\beta^{L(r,\pi^*(\omega))+1}\big)+ \nonumber \\
&\hspace{0.8in}(1-\beta)^2
Q^{L(r,\pi^*(\omega))}(r)\beta^{L(r, \pi^*(\omega))},\nonumber\\
\Lambda_i&=(1{-}\beta)
\beta^{L(r, \pi^*(\omega))}\big[(1-\beta p)R(Q^{L(r,\pi^*(\omega))}(r)){+}\nonumber \\
&\hspace{1.5in}\beta
Q^{L(r,\pi^*(\omega))}(r)R(p) \big]. \nonumber
\end{align}

$\bullet$ If $\pi_i < r_i$, the index value $W_i(\pi_i)$ is given in~(\ref{eq:W2}).
\vspace{10pt}

\noindent Case (2). Negatively correlated channel ($p_i\leq r_i$.)\\

$\bullet$ If $\pi_i \geq r_i$, we have $W_i(\pi_i)=R_i(\pi_i).$

$\bullet$ If $Q_i(p_i) {\leq} \pi_i {<} r_i$, then
\setcounter{equation}{22}

\begin{align}
W_i(\pi_i){=}\frac{(1{-}\beta)[1{-}\beta(1{-}r_i)]R(\pi_i){+}\beta(1{-}\beta)(1{-}\pi_i)R(r_i)}{[1-\beta \pi_i][1-\beta(1-r_i)]-\beta^2(1-\pi_i)r_i}. \nonumber
\end{align}

$\bullet$ If $\pi_i^0 {\leq} \pi_i {<} Q_i(p_i)$,  the index value is expressed as

\begin{align}
W_i(\pi_i){=}\frac{(1{-}\beta)R(\pi_i)\Delta_i {+}\beta (1{-}\beta) \pi_i \Omega_i {+}\beta(1{-}\beta)(1{-}\pi_i) \Upsilon_i}{\Delta_i{-}\beta(1{-}\beta)(1{-}\beta(1{-}r_i))\pi_i{-}(1{-}\beta)\beta^2 r_i (1{-}\pi_i)}, \nonumber
\end{align}
where
\begin{align}
\Delta_i&{=}\big(1{-}\beta(1{-}r_i)\big)\big(1{-}\beta^2 Q_i(p_i)\big){-}\beta^3 r_i \big(1-Q_i(p_i)\big), \label{eq:Delta}\\
\Omega_i&{=}\beta \big(1{-}\beta(1{-}r_i)\big) R_i(Q_i(p_i)){+}\beta^2 \big(1{-}Q_i(p_i)\big) R_i(r_i), \label{eq:Omega}\\
\Upsilon_i&=\beta^2 r_i R_i(Q_i(p_i))+(1-\beta^2 Q_i(p_i))R_i(r_i). \label{eq:Upsilon}
\end{align}

$\bullet$ If $p_i\leq \pi_i<\pi_i^0$, $W_i(\pi_i)$ is given in~(\ref{eq:W3}),
where $\Delta_i$, $\Omega_i$ and $\Upsilon_i$ are given by~(\ref{eq:Delta})-(\ref{eq:Upsilon}), respectively.
\vspace{8pt}

$\bullet$ If $\pi_i<p_i$,  the index value $W_i(\pi_i)$ is given in equation~(\ref{eq:W4}).

\end{document}